\documentclass[journal, 12pt, draftcls,onecolumn]{IEEEtran}
\usepackage{graphicx}
\usepackage{subfigure}
\usepackage{float}
\usepackage{amsmath}
\usepackage{epstopdf}
\usepackage{cases}
\usepackage{color}
\usepackage{algorithm}  
\usepackage{booktabs}
\usepackage{algorithmic}
\usepackage{amsmath}  
\usepackage{verbatim}
\usepackage{cite}

\usepackage[textsize=small]{todonotes}

\usepackage{lipsum}
\makeatletter
\newenvironment{breakablealgorithm}
{
	\begin{center}
		\refstepcounter{algorithm}
		\hrule height.8pt depth0pt \kern2pt
		\renewcommand{\caption}[2][\relax]{
			{\raggedright\textbf{\ALG@name~\thealgorithm} ##2\par}%
			\ifx\relax##1\relax 
			\addcontentsline{loa}{algorithm}{\protect\numberline{\thealgorithm}##2}%
			\else 
			\addcontentsline{loa}{algorithm}{\protect\numberline{\thealgorithm}##1}%
			\fi
			\kern2pt\hrule\kern2pt
		}
	}{
	\kern2pt\hrule\relax
\end{center}
}
\makeatother

\ifCLASSINFOpdf
\else
\fi
\usepackage{caption}
\usepackage{mathtools}

\begin{document}

\title{ISAC-NET: Model-driven Deep Learning for Integrated Passive Sensing and Communication}

\author{
	Wangjun Jiang,~\IEEEmembership{Student Member,~IEEE,} 
	Dingyou Ma,~\IEEEmembership{} \\
	Zhiqing Wei,~\IEEEmembership{ Member,~IEEE,} 
	Zhiyong Feng,~\IEEEmembership{Senior Member,~IEEE,} \\
	Ping Zhang,~\IEEEmembership{Fellow,~IEEE}
	\\
	\thanks{This work is supported in part by the National Key Research and Development Program under Grant 2020YFA0711302, and in part by the BUPT Excellent Ph.D. Students Foundation under Grant CX2022207. \emph\{{Corresponding author: Zhiyong Feng and Zhiqing Wei.\}}
		
	W. Jiang, D. Ma, Z. Wei, and Z. Feng are with the School of Information and Communication Engineering, Beijing University of Posts and Telecommunications, and also with the Key Laboratory of Universal Wireless Communications, Ministry of Education, Beijing 100876, China (email: \{jiangwangjun, dingyouma, weizhiqing, fengzy\}@bupt.edu.cn). 
		
		P. Zhang is with the School of Information and Communication Engineering, Beijing University of Posts and Telecommunications, and also with the State
		Key Laboratory of Networking and Switching Technology, Beijing 100876,
		China (email: pzhang@bupt.edu.cn). 
		}
}

\maketitle

\begin{abstract}
Recent advances in wireless communication with the enormous demands of sensing ability have given
rise to the integrated sensing and communication (ISAC) technology, among which passive sensing plays an important role. The main challenge of passive sensing is how to achieve high sensing performance in the condition of communication demodulation errors.
In this paper, we propose an ISAC network (ISAC-NET) that combines passive sensing with communication signal detection by using model-driven deep learning (DL).
Dissimilar to existing passive sensing algorithms that first demodulate the transmitted symbols and then obtain passive sensing results from the demodulated symbols,
ISAC-NET obtains passive sensing results and communication demodulated symbols simultaneously. 
Different from the data-driven DL method, we adopt the block-by-block signal processing method that divides the ISAC-NET into the passive sensing module, signal detection module and channel reconstruction module.
From the simulation results, ISAC-NET obtains better communication performance than the traditional signal demodulation algorithm, which is close to OAMP-Net2. 
Compared to the 2D-DFT algorithm, ISAC-NET demonstrates significantly enhanced sensing performance. 
In summary, ISAC-NET is a promising tool for passive sensing and communication in wireless communications.
\end{abstract}


\IEEEpeerreviewmaketitle

\section{Introduction}\label{sec:intro}
 
 {\color{black}
 	Integrated sensing and communication (ISAC) systems have been researched independently for many years. Recently, there has been a growing interest in integrating both functions into a single system. This involves sharing the hardware, signal processing modules, and transmitting signals. There are many potential benefits to the close integration of communication and sensing functions, such as improving spectral efficiency, reducing device size, cost, and power consumption, and enhancing the performance of both functions \cite{[ISAC-Fang]}. To achieve efficient integration, advanced signal processing techniques play an important role in the design of transmitting signals and receiver processing \cite{[S_0]}.

  Communication is the process of receiving the ISAC signal at the receiver and using the channel state information (CSI) to demodulate the transmitted symbols.
  The {\color{black} transmitter} and receiver of the communication are separated.
  In contrast, after receiving the ISAC signal, the receiver of sensing uses the known transmitted signal to obtain the target parameters including the velocity and distance of targets.
  Sensing can be divided into two categories: active sensing and passive sensing \cite{[PMC_3]}.
  For active sensing, the transmitter and receiver are co-located, which means that the receiver knows the transmitted signal in advance.
  For this reason, the active sensing receiver can recover the target parameters by using the transmitted signal and the received echoes. This is due to the fact that the transmitted signal interacts with the target and generates echoes that contain valuable information about the target, which can be processed by the receiver to extract the target parameters.
 
  Compared to active sensing, passive sensing has advantages in concealment, efficiency and security \cite{[PMN]}.
  {\color{black} In addition, the signal interaction process of passive sensing is very similar to that of communication, which can be achieved through direct reuse of existing communication devices. Therefore, this paper focuses on integrated passive sensing and communication.
  }
  In passive sensing, the transmitter and receiver are separated, similar to communication. As the receiver does not have prior knowledge of the transmitted signal, it cannot recover the target parameters solely from the received signal. 
  To address the passive sensing problem, a potential solution is to use pilot signal based passive sensing. 
  In pilot signal based passive sensing, a known pilot signal is transmitted, and the receiver can use the received pilot signal to estimate the target parameters by the sensing algorithm \cite{[S_35],[S_37],[S_40],[S_41],[AOA_9],[S_38]}. 
  Two-dimensional discrete Fourier transform (2D-DFT) is a typical sensing algorithm with low complexity and easy implementation \cite{[OFDM]}. 
  
  Nevertheless, the time-frequency resources of pilot signal are limited, which cannot meet the needs of high-performance target sensing.
  In comparison with the pilot signal, the data signal has abundant time-frequency resources.
  Hence, data signal based passive sensing has gradually become a research hotspot. 
  Distinguished from the pilot signal, the receiver cannot obtain the information of data signal in advance.
  Therefore, how to achieve high passive sensing performance in the condition of communication demodulation errors is a major challenge in data signal based passive sensing.
  For this problem, Berger {\it et al.} proposes a signal processing algorithm for passive radar using orthogonal frequency division multiplexing (OFDM) signals \cite{[PS_2010], [Li]}. By extracting the direct signal with strong power, the algorithm demodulated the communication information, and then extracted the multipath echo signal reflected by the target. According to the demodulated communication information, the traditional sensing algorithm \cite{[OFDM]} can be used to realize the target parameter estimation. 
  Essentially, the algorithm proposed in \cite{[Li]} treats communication and sensing as separate and independent processes, and fails to leverage the potential performance benefits that can be achieved by integrating the two functions. 
  Based on the above analysis, the communication demodulation information and passive sensing information can support each other in the integrated passive sensing and communication system. Specifically, the target delay and Doppler information obtained from the result of sensing are helpful to the demodulation of communication information, and the demodulation of communication information can also stimulate the target parameter estimation. 
  Consequently, making full use of the mutual auxiliary relationship between communication and sensing functions is one of the key technologies to improve both communication and sensing performance.
}
 
 Recognize this fact, DL based signal processing algorithms are gradually gaining attention.
 The potential applications of DL in wireless communications are discussed in \cite{[COM_1],[COM_2],[COM_3]}. The data-driven DL method in \cite{[COM_4]} adopts a fully connected deep neural network (FC-DNN) and replaces all modules at the OFDM receiver. The above-mentioned literature regards the traditional OFDM receiver as a black box. Nonetheless, it does not exploit the expert knowledge in wireless communications, which in turn renders the FC-DNN based receiver unexplainable and unpredictable.
 To address the above issues, the model-driven DL approach can be used instead. This approach involves explicitly incorporating domain knowledge and prior information into the model, which can improve its robustness and accuracy under error conditions \cite{[XW_model_driven]}.
 A general model-driven DL architecture, called ComNet \cite{[COM_5]}, combines DL with expert knowledge in wireless communications, which can to replace the conventional or FC-DNN OFDM receiver \cite{[COM_4]}. 
 To further promote the performance of ComNet, the improved ComNet based on expectation propagation (IComNet-EP) and orthogonal approximate message passing network (OAMP-Net2) are proposed in \cite{[COM_6], [Main]}, which have comparative complexity with the ComNet but achieve better robustness for the communication signal detection.
 However, these works do not use ISAC technology, and thus do not provide an effective way for the trade off between communication demodulation and passive sensing.
 {\color{black}
 	It is well-known that sensing and communication can benefit each other based on ISAC. 
 	However, most existing ISAC signal processing algorithms treat sensing and communication demodulation processing algorithms as two independent processes, and do not involve joint processing algorithms to achieve communication and sensing results at the same time \cite{[jwj_1],[jwj_3],[jwj_2]}. 
 	For integrated active sensing and communication, sensing and communication demodulation are processed on different nodes, and the discrete ISAC signal processing algorithm can meet the requirements of this scenario. 
 	However, for integrated passive sensing and communication, sensing and communication demodulation are processed at the same node, and the error of communication demodulation and sensing results will affect each other. In this case, the traditional discrete design of ISAC signal processing algorithm cannot meet the requirements of this scenario.
 }
 {\color{black}
 In summary, the main challenges of the integrated passive sensing and communication can be listed as follow:
\begin{itemize}
	\item How to combine pilot signal and data signal to realize passive sensing for improving sensing performance.
	\item How to optimize the passive sensing and communication demodulation processes to achieve mutual performance gain.
	\item How to break the performance limits of traditional passive perception algorithms through DL.
\end{itemize}
}

 Against the above backdrop, this paper proposes an ISAC network (ISAC-NET) to achieve communication and sensing at the same time. 
 {\color{black}
 The model-driven based on ISAC-NET combines the conventional alternating optimization and deep unfolding into a neural network.
 Compared with traditional data-driven DL networks, the key challenge of model-driven DL networks is how to set network architecture modules and learning parameters reasonably.
 The ISAC-NET considers the key steps of ISAC signal processing, and designs three modules, including the passive sensing module, signal detection module and channel reconstruction module. 
}In the ISAC-NET, communication data demodulation and target parameter estimation are interlaced and interactively optimized to achieve the optimal performance of both functions. 
 The major contributions of this paper are summarized as follows.

{
\begin{enumerate}
\item For integrated passive sensing and communication systems, we propose the ISAC-NET that combines passive sensing with communication demodulation signal processing by using model-driven DL. Different from existing passive sensing algorithms that first demodulate the transmitted symbols and then obtain passive sensing results by using the demodulated symbols, ISAC-NET obtains passive sensing results and communication demodulated symbols simultaneously. Dissimilar to the data-driven DL method, we adopt the block-by-block signal processing method to improve the robustness and accuracy under error conditions. The method divides ISAC-NET into the passive sensing module, signal detection module and channel reconstruction module. 
\item Based on the ISAC-NET proposed in this paper, we have implemented the scheme of passive sensing joint using the data signal and plot signal, which significantly improves the sensing performance. Compared with the traditional passive sensing based on pilot signals, ISAC-NET adopts more time and frequency resources for sensing, thus obtains better passive sensing accuracy.

 {\color{black}
\item We propose the ISAC signal processing algorithm that alternates target parameter estimation and channel reconstruction to realize multi-target sensing. The ISAC signal processing algorithm contains the improved denoising-based approximate message passing (D-AMP) algorithm, the 2D-DFT passive sensing algorithm and the ISAC channel reconstruction algorithm, which will be introduced in $\mathrm{Section\ \ref{sec:ISAC-NET-1}}$. This method has certain improvement in the performance of multi-target sensing, which is verified in the subsequent simulation analysis.
}

\item We analyze and evaluate the performance improvements of the ISAC-NET in range, velocity estimation, and communication signal detection through simulation. The simulation results verify that the ISAC-NET achieves better communication and passive sensing performance, especially in multi-targets detection scenario. In terms of communication, ISAC-NET obtains better communication performance than the traditional signal demodulation algorithm, which is close to OAMP-Net2 \cite{[Main]}. Compared to the 2D-DFT sensing algorithm, ISAC-NET demonstrates significantly enhanced sensing performance and approaches the perfect sensing performance, that is, the sensing performance without communication demodulation errors.
 \end{enumerate}
}

The remaining parts of this paper are organized as follows.
$\mathrm{Section\ \ref{sec:system_model}}$ describes the system model of integrated passive sensing and communication. 
The design of ISAC-NET is introduced in $\mathrm{Section\ \ref{sec:ISAC-NET-1}}$ and $\mathrm{\ref{sec:ISAC-NET-2}}$. 
The performance improvements in sensing and communication are analyzed and simulated in $\mathrm{Section\ \ref{sec:Simulation}}$. $\mathrm{Section\ \ref{sec:Conclusion}}$ concludes the paper. 

Notations: The symbols used in this paper are described as follows. Vectors and matrices are denoted by boldface small and capital letters. The transpose, complex conjugate, Hermite, inverse, and pseudo-inverse of the matrix ${\bf{A}}$ are denoted by ${{\bf{A}}^{\rm T}}$, ${{\bf{A}}^*}$, ${{\bf{A}}^{\rm H}}$, ${{\bf{A}}^{ - 1}}$ and ${{\bf{A}}^\dag}$, respectively. ${\rm{diag}}(\bf{x})$ is the operation that generates a diagonal matrix with the diagonal elements to be the elements of $\bf x$. $ \otimes $ is the Kronecker product operator. $ \odot $ is the Hadamard product operator. 
{
${{\bf A}}({m,n})$ denotes the element in row $m$, column $n$ of the matrix ${{\bf A}}$. $\mathcal{S}({\bf A}_1,  {\bf A}_2)$ denotes the set of ${\bf A}_1$ and ${\bf A}_2$. $P({\bf A})$ denotes the probability of $\bf A$. For the sake of convenience, key parameters and abbreviations used in this paper are given in $\mathrm{TABLE\ \ref{label:abbreviations}}$.
}

\begin{table}[ht]
	\centering
	\caption{Key Parameters and Abbreviations}
	\label{label:abbreviations}
	\begin{tabular}{l|l|l|l}
		\hline \hline
		Abbreviation & Description & Abbreviation & Description \\ \hline
		DL & Deep learning & ISAC & Integrated sensing and communication \\ \hline
		LoS & Line of sight & NLoS & Non-LoS \\ \hline
		AoA & Angle of arrival & AoD & Angle of departure \\ \hline
		CSI & Channel state information & 2D-DFT & Two-dimensional
		discrete Fourier transform \\ \hline
		OFDM & Orthogonal frequency division multiplexing & QAM & Quadrature amplitude modulation \\ \hline
		ComNet & Communication network & IComNet-EP & Improved ComNet based on expectation propagation \\ \hline
		OAMP & Orthogonal approximate message passing & OAMP-Net2 & OAMP network  \\ \hline
		BER & Bit error rate & SER & Symbol error rate \\ \hline
		SNR & Signal-to-noise ratio & NMSE & Normalized mean squared error \\ \hline
		FC-DNN & Fully connected deep neural network & ISAC-NET & ISAC network \\ \hline
	\end{tabular}
\end{table}

\section{System Model}\label{sec:system_model}

\subsection{Integrated Passive Sensing and Communication}\label{sec:system_model-1}

\begin{figure}[ht]
	\includegraphics[scale=0.8]{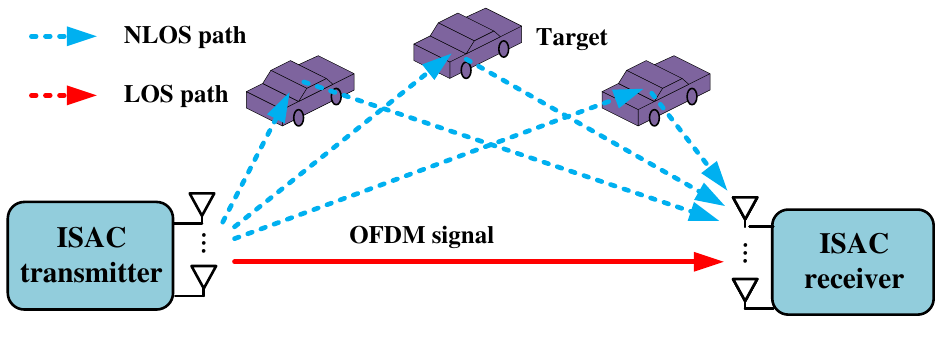}
	\centering
	\caption{Passive sensing and communication.}
	\label{fig:ISAC_system_model}
\end{figure}

\begin{figure}[ht]
	\includegraphics[scale=0.8]{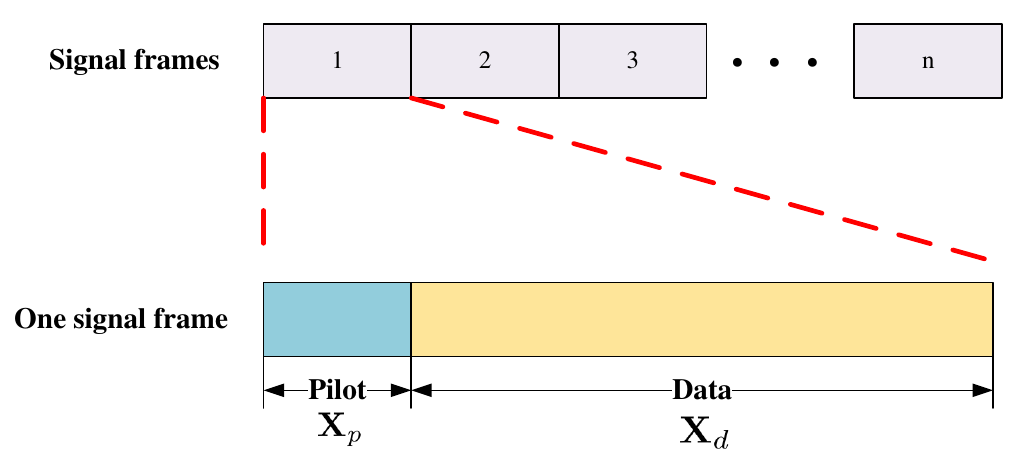}
	\centering
	\caption{Signal frame of passive sensing and communication.}
	\label{fig:Signal_frame}
\end{figure}

We consider the system model of passive sensing and communication, as shown in $\mathrm{Fig.\ \ref{fig:ISAC_system_model}}$. The ISAC receiver receives the multipath ISAC signals from the ISAC transmitter.
{\color{black}
 As shown in $\mathrm{Fig.\ \ref{fig:ISAC_system_model}}$, the line of sight (LoS) path signal is not reflected by the target and directly reaches the receiver. It only carries the communication information and does not carry the target sensing information.
Non-LoS (NLoS) signals are target echoes, which can be used not only for demodulation of communication information, but also for passive sensing of the target. 
Nonetheless, noise can cause interference to communication and passive sensing.
}

\subsection{Transmitted Signal Model}\label{sec:system_model-Tx}
As the main signal model of 5G, OFDM is adopted as the transmitted signal model of ISAC system, which can be expressed as \cite{[OFDM]}
\begin{equation}\label{equ:OFDM_signal}
	x(t) = \sum\limits_{m = 0}^{M - 1} {\sum\limits_{n = 0}^{N - 1} {s\left(mN + n\right)} \cdot e^{j2\pi{(f_c + n \Delta f)}t} } {\mathrm {rect}} \left(\frac{{t - mT}}{T}\right),
\end{equation}
where $M$ is the number of OFDM symbols, $N$ is the number of subcarriers, $m$ is the OFDM symbol index, $n$ is the index of subcarrier, $s(mN + n)$ is the complex modulation symbol, ${\mathrm {rect}} (\cdot)$ is the rectangle function, $f_c$ is the carrier frequency, ${\Delta f}$ denotes subcarrier spacing. The duration time of each OFDM symbol $T$ contains the elementary symbol duration ${T_{p}}$ and the guard interval ${T_c}$. If ${T_c}$ is larger than the maximum multipath delay, the inter-symbol interference can be eliminated. 

\subsection{Received Signal Model}\label{sec:system_model-2}

Assume that the ISAC transmitter and receiver has a limited number of $N_t$ and $N_r$ antennas. As $\mathrm{Fig.\ \ref{fig:Signal_frame}}$ shows, the ISAC signal ${\bf X}$ for passive sensing and communication includes the pilot signal ${\bf X}_p$ and data signal ${\bf X}_d$. 
{\color{black}
Pilot signals generally send a known reference signal at the receiving end, including CSI, position reference signal (PRS), etc. In the integrated communication and passive sensing system, pilot signals can be used for channel estimation and initial passive sensing.
On the one hand, the longer the pilot signal, the higher the accuracy of the initial passive sensing, and the better sensing performance of ISAC-NET can be obtained.
However, the pilot signal does not carry communication information, long pilot signal will reduce the communication capacity.
In the field of communication, the length of the pilot signal has been studied widely, and different pilot parameter setting standards have been proposed for different scenarios.
According to 5G NR standard \cite{[5G_signal]}, typically, 1 or 2 OFDM symbols are selected as the pilot signal among the 14 OFDM symbols in each time slot of the OFDM signal.
without loss of generality, this paper adopts a typical pilot length ratio setting, 1/14, to conduct subsequent analysis and simulation experiments.
}

Then, the transmitted and received signal matrix from the $i$-th transmitting antenna to the $k$-th receiving antenna can be expressed as 
\begin{equation}\label{equ:OFDM_receive_all_1}
	{{\bf X}'}_{i \rightarrow k} = \begin{bmatrix}  
		x_{i \rightarrow k}(0,0)& \cdots  & x_{i \rightarrow k}(0,N-1) \\  
		x_{i \rightarrow k}(1,0)& \cdots  & x_{i \rightarrow k}(1,N-1) \\  
		\vdots & \ddots & \vdots \\  
		x_{i \rightarrow k}(M-1,0)& \cdots  & x_{i \rightarrow k}(M-1,N-1) 
	\end{bmatrix},
\end{equation}
\begin{equation}\label{equ:OFDM_receive_all_2}
	{{\bf Y}'}_{i \rightarrow k} = \begin{bmatrix}  
		y_{i \rightarrow k}(0,0)& \cdots  & y_{i \rightarrow k}(0,N-1) \\  
		y_{i \rightarrow k}(1,0)& \cdots  & y_{i \rightarrow k}(1,N-1) \\  
		\vdots & \ddots & \vdots \\  
		y_{i \rightarrow k}(M-1,0)& \cdots  & y_{i \rightarrow k}(M-1,N-1) 
	\end{bmatrix},
\end{equation}
where the $n$-th subcarrier $m$-th OFDM symbol of $	{{\bf Y}'}_{i \rightarrow k}$ can be expressed as \cite{[Main]}
\begin{equation}\label{equ:ISAC_new_1_1}
	\begin{aligned}
	{{y}}_{i \rightarrow k}({m,n}) = \sum_{l=0}^{L_p -1} \widetilde{f}_{l}  \widetilde{F}_{l} & \cdot e^{{- j 2 \pi ({f_c + n \Delta f}) \tau_l}} \cdot e^{{j 2 \pi f_{d,l} mT}} 
	 \cdot e^{j2\pi \frac{d_a}{\lambda} (i-1) {\mathrm {sin}}(\theta_t^l)} 
	 \cdot e^{j2\pi \frac{d_a}{\lambda} (k-1) {\mathrm {sin}}(\theta_r^l)} \\
	& \cdot {{x}}_{i \rightarrow k}({m,n})  + 	{{z}}_{i \rightarrow k}({m,n}), 
	\end{aligned}
\end{equation}
\begin{equation}\label{equ:ISAC_new_1_2}
	\begin{aligned}
	\tau_l = \frac{r_l}{c},
	\end{aligned}
\end{equation}
\begin{equation}\label{equ:ISAC_new_1_3}
	\begin{aligned}
		f_{d,l} = \frac{v_l f_c}{c},
	\end{aligned}
\end{equation}
where $c$ is the speed of light, $\lambda$ is the wavelength of ISAC signal, 
$d_a = \frac{\lambda}{2}$ is the distance between the adjacent elements of the transmit and receive antenna arrays, 
$\theta_t^l$ and $\theta_r^l$ are the angle of departure (AoD) and angle of arrival (AoA) of the $\it l$-th path,
$\tau_l$ and $f_{d,l}$ are the time delay and Doppler frequency shift of the $l$-th path, respectively,
The small scale fading factor $\widetilde{f}_{l}$ obeys the Weibull distribution \cite{[Weibull_181]}, $\widetilde{F}_{l} \approx \frac{\sigma \lambda^2}{(4\pi)^3 {(r_{1,l}r_{2,l})^2}}$ is the large scale fading factor, $r_{1,l}$ and $r_{2,l}$ are the range from the transmitter to the target and the target to the receiver of the $l$-th path respectively.
The range of the $l$-th path $r_{l}$ can be derived as $r_{1,l} + r_{2,l}$.
$\sigma$ is the radar cross section (RCS) of targets, $L_p$ is the number of multiple paths, ${\bf X}_{i \rightarrow k}' \in {{\mathcal C}^{M \times N}}$ is the transmitted OFDM signal, ${\bf Y}_{i \rightarrow k}' \in {{\mathcal C}^{M \times N}}$ is the received OFDM signal, ${\bf Z}_{i \rightarrow k}' \in {{\mathcal C}^{M \times N}} $ is the additive white Gaussian noise (AWGN) matrix.
Then, in the multiple-input multiple-output (MIMO) system, 
the $m$-th OFDM symbol of the received signal with the $n$-th subcarrier is composed into a vector ${\bf y}({m,n})$, which is expressed as 
\begin{equation}\label{equ:ISAC_new_2}
	{\bf y}({m,n})  = \sum_{l=0}^{L_p -1} \tilde{{\bf H}}^l ({m,n}) {\bf x}({m,n}) + {\bf z}({m,n}), 
\end{equation}
where ${\bf y}({m,n}) \in {{\mathcal C}^{N_r \times 1}}$ is the $m$-th symbol and $n$-th subcarrier OFDM received signal, ${\bf x}({m,n}) \in {{\mathcal C}^{N_t \times 1}}$ is the transmitted signal of the $m$-th symbol and the $n$-th subcarrier. ${\bf z}({m,n}) \in {{\mathcal C}^{N_r \times 1}}$ is the AWGN vector. The channel matrix of the $l$-th path $\tilde{{\bf H}}^l ({m,n}) \in {{\mathcal C}^{N_r \times N_t}}$ can be expressed as 
\begin{equation}\label{equ:ISAC_C_1}
	\begin{aligned}
	\tilde{{\bf H}}^l ({m,n}) = \widetilde{f}_{l}  \widetilde{F}_{l} \cdot e^{{- j 2 \pi ({f_c + n \Delta f}) \tau_l}} \cdot e^{{j 2 \pi f_{d,l} mT}} 
	\cdot {\bf d}_r(\theta_r^l) {\bf d}_t^H(\theta_t^l),
	\end{aligned}
\end{equation}
where ${\bf d}_t$ and ${\bf d}_r$ are the transmitting and receiving steering vector, which can be expressed as
\begin{equation}\label{equ:ISAC_C_2}
	\begin{aligned}
		{\bf d}_t(\theta_t^l) &= \frac{1}{\sqrt{N_t}} \left [1,..., e^{j 2\pi \frac{d_a}{\lambda} (N_t-1) {\rm sin}(\theta_t^l) } \right] \\
		{\bf d}_r(\theta_r^l) &= \frac{1}{\sqrt{N_r}} \left[1,..., e^{j 2\pi \frac{d_a}{\lambda} (N_r-1) {\rm sin}(\theta_r^l) } \right]
	\end{aligned}.
\end{equation}
Note that \eqref{equ:ISAC_new_2} can be expressed into a matrix form as
\begin{equation}\label{equ:RSM_1}
	\begin{aligned}
		{\bf Y} & = \sum_{l=0}^{L_p -1} {\bf H}^l_a {\bf X} {\bf H}^l_{\varphi} + {\bf Z}, 
	\end{aligned}
\end{equation}
\begin{equation}\label{equ:ISAC_new_4_1}
	\begin{aligned}
		{\bf H}^l = \mathcal{S} ( {\bf H}^l_a,  {\bf H}^l_{\varphi}  ),
	\end{aligned}
\end{equation}
where 
\begin{equation}\label{equ:ISAC_new_3}
	\begin{aligned}
		{\bf Y} &= \left[{\bf y}_1, {\bf y}_2, \cdots , {\bf y}_{N_r}\right]^T \in {{\mathcal C}^{N_r \times MN}}, \\
		{\bf X} &= \left[{\bf x}_1, {\bf x}_2, \cdots , {\bf x}_{N_t}\right]^T \in {{\mathcal C}^{N_t \times MN}}, \\
		{\bf Z} &= \left[{\bf z}_1, {\bf z}_2, \cdots , {\bf z}_{N_r}\right]^T \in {{\mathcal C}^{N_r \times MN}},
	\end{aligned}
\end{equation}
with
\begin{equation}\label{equ:ISAC_new_4}
	\begin{aligned}
		{\bf y}_i &= \mathrm{vec} ({\bf Y}_i') \in {{\mathcal C}^{MN \times 1}}, \\
		{\bf x}_i &= \mathrm{vec} ({\bf X}_i') \in {{\mathcal C}^{MN \times 1}}, \\
		{\bf z}_i &= \mathrm{vec} ({\bf Z}_i') \in {{\mathcal C}^{MN \times 1}},
	\end{aligned}
\end{equation}
where ${\bf x}_i$ is the transmitted signal from the $i$-th antenna, ${\bf y}_i$ is the received signal from the $i$-th antenna, ${\bf z}_i$ is the noise signal from the $i$-th antenna.
The channel matrix of the $l$-th path ${\bf H}^l$ consists of two parts: channel amplitude information matrix ${\bf H}^l_a \in {{\mathcal C}^{N_r \times N_t}}$ and channel phase information matrix ${\bf H}^l_{\varphi} \in {{\mathcal C}^{MN \times MN}}$
\begin{equation}\label{equ:ISAC_new_4_2}
	\begin{aligned}
		{\bf H}^l_a =  \widetilde{f}_{l}  \widetilde{F}_{l} \cdot {\bf d}_r(\theta_r^l) {\bf d}_t^H(\theta_t^l) ,
	\end{aligned}
\end{equation}
{
\begin{equation}\label{equ:ISAC_new_4_3}
	\begin{aligned}
		{\bf H}^l_{\varphi} &=  &{\rm{diag}}\left(   {h}^l_{\varphi}(0,0), \cdots,{h}^l_{\varphi}(m,n), 
		\cdots, {h}^l_{\varphi}(M-1,N-1)  \right) ,
	\end{aligned}
\end{equation}
with
\begin{equation}\label{equ:ISAC_new_4_4}
	\begin{aligned}
		{{h}}^l_{\varphi}(m,n) =  e^{{- j 2 \pi ({f_c + n \Delta f}) \tau_l}} \cdot e^{{j 2 \pi f_{d,l} mT}} .
	\end{aligned}
\end{equation}
}
\begin{figure}[ht]
	\includegraphics[scale=0.4]{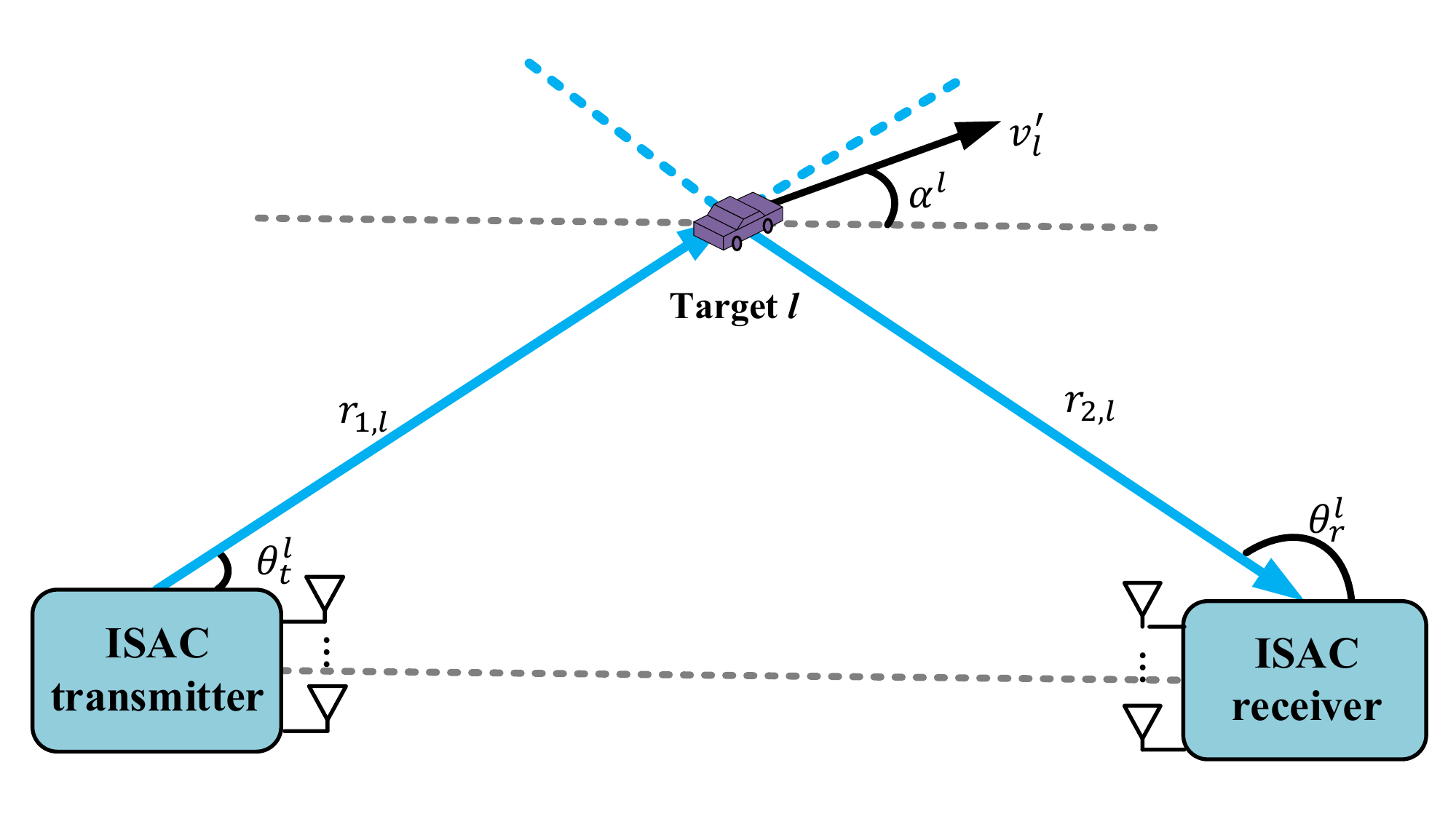}
	\centering
	\caption{Delay and doppler frequency shift for passive sensing.}
	\label{fig:PS_vs_AS}
\end{figure}
As $\mathrm{Fig.\ \ref{fig:PS_vs_AS}}$ shows, the delay $\tau_l$ and doppler frequency shift ${f}_{d,l}$ of the $l$-th target can be expressed as  
\begin{equation}\label{equ:ISAC_NET_3-1-1}
	\begin{aligned}
		\tau_l &= \frac{r_l}{c} = \frac{r_{1,l} + r_{2,l}}{c}, 
	\end{aligned}
\end{equation}
\begin{equation}\label{equ:ISAC_NET_3-1-2}
	\begin{aligned}
		{f}_{d,l} &= \frac{v_l}{\lambda} =  -\frac {{v_l}'}{c} f_c \cdot {\rm cos} \left(\frac{\theta_r^l - \theta_t^l}{2}\right) \cdot {\rm cos} \left(\alpha^l - \frac{\theta_r^l + \theta_t^l}{2}\right),
	\end{aligned}
\end{equation} 
where $\alpha^l$ is the angle of the $l$-th target velocity in the horizontal direction; ${v_l}'$ is the absolute velocity of the $l$-th target; $v_l$ is the radial velocity of the $l$-th target. 
Passive sensing can estimate the time delay and Doppler shift to get the $r_{l}$ and $v_l$ of the $l$-th target.
The proof of \eqref{equ:ISAC_NET_3-1-2} is provided in $\mathrm{Appendix\ \ref{app:A}}$.

\section{ISAC Signal Processing}\label{sec:ISAC-NET-1}

According to the above analysis, the signal processing of integrated passive sensing and communication can be expressed as 
\begin{equation}\label{equ:ISAC_NET_1}
	\begin{aligned}
		{\mathcal{P}} : & \min_{{{\bf X}},[ {\bf r}, {\bf v}]}  \quad \begin{Vmatrix}
			{\bf Y} - \sum_{l=0}^{L_p -1} {\bf H}^l_a {\bf X} {\bf H}^l_{\varphi}
		\end{Vmatrix}_2 \\
		& {\rm {s.t.}} \qquad \left[ {\bf r}, {\bf v}\right] = {\mathcal F}_{s}({\bf H}^l)
		\\
		&\qquad \qquad {\bf H}^l = \mathcal{S} ({\bf H}^l_a,  {\bf H}^l_{\varphi} )    \\
		&\qquad \qquad \quad = {\mathcal F}_{h} ([ {\bf r}, {\bf v}])
	\end{aligned},
\end{equation}
where
\begin{equation}\label{equ:r_v}
	\begin{aligned}
		{\bf r} &=  \left[ r_0, r_1, \cdots, r_l,\cdots, r_{L_p-1}   \right]^T \\
		{\bf v} &= \left[ v_0, v_1, \cdots, v_l,\cdots, v_{L_p-1}   \right]^T
	\end{aligned},
\end{equation}
are the estimated range and velocity vectors of targets,
${ {\bf X}}$ is the detected transmitted signal.
{\color{black}
As shown in \eqref{equ:ISAC_NET_1}, the objective of the optimization problem $\min_{{{\bf X}},[ {\bf r}, {\bf v}]}   \begin{Vmatrix} {\bf Y} - \sum_{l=0}^{L_p -1} {\bf H}^l_a {\bf X} {\bf H}^l_{\varphi} \end{Vmatrix}_2$ is to minimize the error between the reconstructed received signal and the real received signal $\bf Y$. 
To achieve the objective, there are two constraints: accurately demodulate the transmitted signal $\bf X$, accurately estimate environmental targets and reconstruct channels.
}

{
The passive sensing function ${\mathcal F}_{s} (\cdot)$ is the 2D-DFT algorithm based sensing signal processing algorithm \cite{[OFDM],[S_0]}, which will be introduced in $\mathrm{Section\ \ref{sec:ISAC-NET-1-2}}$.
The ISAC channel reconstruction function ${\mathcal F}_{h}  (\cdot)$ will be introduced in $\mathrm{Section\ \ref{sec:ISAC-NET-1-3}}$.

According to \eqref{equ:ISAC_NET_1}, 
the signal processing of integrated passive sensing and communication is to simultaneously optimize communication demodulation and target parameter estimation to achieve optimal communication and sensing performance. 
The problem solved by signal processing is a {\color{black} multi-parameter} optimization problem, which includes the detected transmitted signal $\bf X$ and radar sensing results $[\bf r, v]$.
The iteration algorithm is a typical algorithm to solve the {\color{black} multi-parameter} optimization problem \cite{[CI]}.
Therefore, we propose an ISAC signal processing algorithm based on iteration, which consists of the following iterations
\begin{equation}\label{equ:ISAC_NET_2}
	\begin{aligned}
		{\bf X}^{k+1} &=  {\mathcal F}_1 \left(\bf X ^{\it k}, {\bf H}^{\it k} \right) \\ 
		\left[{\bf r}^{k+1},{\bf v}^{t+1}\right] &=  {\mathcal F}_2 \left({\bf X}^{k+1}, \bf Y \right) \\
		{\bf H}^{k+1} &=  {\mathcal F}_3 \left({\bf r}^{k+1},{\bf v}^{k+1} \right) 
	\end{aligned},
\end{equation}
where $k$ is the index of iterations, $\bf H$ represents the channel matrix set of $L_p$ paths, ${\mathcal F}_1$, ${\mathcal F}_2$ and ${\mathcal F}_3$ are the optimization function of $\bf X$, $\left[\bf r,\bf v\right]$ and $\bf H$, respectively. 

\begin{table}[ht]
	\centering
	\caption{Iterations in ISAC signal processing}
	\label{label:Iterations}
	\begin{tabular}{l|l|l|l}
		\hline \hline
		 & ${\mathcal F}_1$ & ${\mathcal F}_2$ & ${\mathcal F}_3$ \\ \hline
		Name & Signal detection algorithm & Passive sensing algorithm & Channel reconstruction algorithm \\ \hline
		Specific form & 
		\begin{tabular}[c]{@{}l@{}} 
			$
			{\bf X}_{d}^{k+1} = \eta_1^k {\bf X}_{d}^{k} + \sum_{l=0}^{L_p -1} {\hat {{ {\bf H}}_a^l}}^T {\bf Z}_{d}^{k} {\hat {{ {\bf H}}_{\varphi}^l}}^T.
			$ \\ 
			$
			{\bf Z}_{d}^{k+1} =  \eta_2^k {\bf Z}_{d}^{k} + {\bf Y}_d - \sum_{l=0}^{L_p -1} \hat {\bf H}^l_a {\bf X}_{d}^{k} \hat {\bf H}^l_{\varphi}.
			$ 
		\end{tabular}
		  &
		 \begin{tabular}[c]{@{}l@{}} 
		 	$
		 	{\bf r}^{k+1} = \mathsf{IDFT} \left( {\bf X}^{k+1}, {\bf Y} \right).
		 	$ \\ 
		 	$
		 	{\bf v}^{k+1} = \mathsf{DFT} \left( {\bf X}^{k+1}, {\bf Y} \right).
		 	$ 
		 \end{tabular}
		   & 
		   $
		   {\bf H}^{k+1} = \mathsf {CHA} \left( {\bf r}^{k+1},{\bf v}^{k+1} \right)
		   $
		    \\ \hline
		Details in & $\mathrm{Section\ \ref{sec:ISAC-NET-1-1}}$ & $\mathrm{Section\ \ref{sec:ISAC-NET-1-2}}$ & $\mathrm{Section\ \ref{sec:ISAC-NET-1-3}}$ \\ \hline
	\end{tabular}
\end{table}
}

\subsection{${\mathcal F}_1$: Improved D-AMP Signal Detection Algorithm}\label{sec:ISAC-NET-1-1}
{
	Based on \eqref{equ:RSM_1}, the received data signal can be expressed as
	\begin{equation}\label{equ:ISAC_NET_2_1}
		\begin{aligned}
			{\bf Y}_d = \sum_{l=0}^{L_p -1} {\bf H}^l_a {\bf X}_d {\bf H}^l_{\varphi} + {\bf Z}_d
		\end{aligned},
	\end{equation}
where ${\bf Z}_d$ is the AWGN on data signal. Moreover, ${\bf X}$ consists of ${\bf X}_p$ and ${\bf X}_d$; ${\bf Y}$ consists of ${\bf Y}_p$ and ${\bf Y}_d$; ${\bf Z}$ consists of ${\bf Z}_p$ and ${\bf Z}_d$.
The maximum posterior (MAP) is the typical algorithm for recovering the transmitted data signal ${\bf X}_d$ from the received data signal by maximizing the posterior probability ${\it P} \left({\bf X}_d | {\bf Y}_d, {\hat {\bf H}^l}\right)$. 
However, {\color{black} the computational complexity of MAP is high} due to the large number of matrix inversion operations required. To address this problem, the D-AMP algorithm has been proposed in \cite{[D_AMP]}. 
The D-AMP algorithm decouples the posterior probability ${\it P} \left({\bf X}_d | {\bf Y}_d, {\hat {\bf H}^l}\right)$ into a series of ${\it P} \left({\bf X}_{d}^k | {\bf Y}_d, {\hat {\bf H}^l}\right), \left(k = 1, 2, . . . , K\right)$ in an iterative way, {\color{black} with $K$ being the number of iterations. The estimated channel ${\hat {\bf H}^l}$ and the received data signal ${\bf Y}_d$ can be obtained in advanced.}
Compared with AMP and OAMP mentioned in $\mathrm{Section\ \ref{sec:intro}}$, the D-AMP algorithm can obtain better signal detection performance than AMP and less computational complexity than OAMP.
}
The denoising function is a mathematical operation that removes noise from a signal, which is one of the key factors that determine the performance of D-AMP \cite{[D_AMP]}. There are many related works on the denoising function, such as wavelet domain denoising \cite{[wavelet_D]}. However, these methods need to determine different hyperparameters depending on the particular scenario. 
To solve the problem, we propose the improved D-AMP algorithm in $\mathrm{Algorithm\ \ref{alg:D_AMP}}$, which mainly contains two modules: signal updating \eqref{equ:ISAC_NET_3} and noise updating \eqref{equ:ISAC_NET_4}. 
Different from the conventional D-AMP, the improved D-AMP sets two trade off parameters $\eta_1$ and $\eta_2$, and adopts the DL method to estimate the two parameters, which will be introduced in $\mathrm{Section\ \ref{sec:ISAC-NET-2}}$.
\begin{algorithm} 
	\caption{Improved D-AMP algorithm}  
	\label{alg:D_AMP}  
	\begin{algorithmic} 
		\REQUIRE 
		Received data signal ${\bf Y}_d$, estimated channel matrix ${\hat {\bf H}^l} = \mathcal{S} \left(\hat {\bf H}^l_a, \hat {\bf H}^l_{\varphi} \right)$.
		\STATE \quad 1). Initialize: ${\bf Z}_{d}^1 \gets {\bf 0}$, ${\bf X}_{d}^1 \gets {\bf 0}$.
		\STATE \quad 2). Update ${\bf X}_{d}^{k+1}$: 
		\begin{equation}\label{equ:ISAC_NET_3}
			{\bf X}_{d}^{k+1} = \eta_1^k {\bf X}_{d}^{k} + \sum_{l=0}^{L_p -1} {\hat {{ {\bf H}}_a^l}}^T {\bf Z}_{d}^{k} {\hat {{ {\bf H}}_{\varphi}^l}}^T.
		\end{equation} \\
		\STATE \quad 3). Update ${\bf Z}_{d}^{k+1}$: 
		\begin{equation}\label{equ:ISAC_NET_4}
			{\bf Z}_{d}^{k+1} =  \eta_2^k {\bf Z}_{d}^{k} + {\bf Y}_d - \sum_{l=0}^{L_p -1} \hat {\bf H}^l_a {\bf X}_{d}^{k} \hat {\bf H}^l_{\varphi}.
		\end{equation} \\
		\ENSURE Recovered data signal ${\bf X}_{d}^{K+1}$.
	\end{algorithmic}  
\end{algorithm}

\subsection{${\mathcal F}_2$: 2D-DFT Passive Sensing Algorithm}\label{sec:ISAC-NET-1-2}
{\color{black}
In terms of passive sensing algorithms, 2D-DFT is a typical sensing algorithm used in ISAC OFDM signal, which can achieve relatively high sensing performance with low complexity \cite{[OFDM]}.}
The 2D-FFT sensing algorithm aims to extract the distance and velocity information of targets from the received signal.
Since the sensing signal processing of the received signal of different antennas is the same, without loss of generality, the transmitted and received OFDM signal from the $i$-th transmitting antenna to the $k$-th receiving antenna is adopted for passive sensing.
Based on the signal model mentioned in $\mathrm{Section\ \ref{sec:system_model-Tx}}$ and $\mathrm{\ref{sec:system_model-2}}$, the $n$-th subcarrier $m$-th OFDM symbol of the baseband received signal can be rewritten as \cite{[OFDM]}
\begin{equation}\label{equ:OFDM_receive}
	\begin{aligned}
		y_{i \rightarrow k}(m,n) = \sum_{l=0}^{L_p -1} & A_{i \rightarrow k}(m,n)x_{i \rightarrow k}(m,n)\\
		& \cdot e^{- j2\pi n \Delta f \frac{{{r_l}}}{c} } \cdot e^{ - \frac{j2\pi {\it{m}}T{{v_l}{f_c}}}{c} }\\
		& + z_{i \rightarrow k}(m,n),
	\end{aligned}
\end{equation}
\begin{equation}\label{equ:OFDM_receive-2}
	\begin{aligned}
		A_{i \rightarrow k}(m,n) = \widetilde{f}_{l}  \widetilde{F}_{l} \cdot e^{j2\pi \frac{d_a}{\lambda} (i-1) {\mathrm {sin}}(\theta_t^l)} \cdot e^{j2\pi \frac{d_a}{\lambda} (k-1) {\mathrm {sin}}(\theta_r^l)},
	\end{aligned}
\end{equation}
where $z_{i \rightarrow k}(m,n)$ is the AWGN, $x_{i \rightarrow k}(m,n)$ is the sending modulation signal. The rectangular window function ${\rm rect}(\cdot)$ with time delay can be neglected in baseband processing \cite{[OFDM]}. 
The transmitted signal $x_{i \rightarrow k}(m,n)$ can be obtained by improved D-AMP algorithm introduced in $\mathrm{Section\ \ref{sec:ISAC-NET-1-1}}$.
Then, after removing the transmitted information from the received information symbols by an element-wise complex division, the elements of division matrix expression can be derived as
\begin{equation}\label{equ:OFDM_signal_1}
	\begin{aligned}
		\left({{\bf{S}}_g}\right)_{m,n} &= \frac{{{y}}_{i \rightarrow k} (m,n)}{{{x}}_{i \rightarrow k} (m,n)} \\
		&\approx 
		A_{i \rightarrow k}(m,n){k_r(n)}{k_v(m)}
	\end{aligned},
\end{equation}
where
\begin{equation}\label{equ:OFDM_signal_2}
	{k_r(n)} = e^{- j2\pi {n \Delta f}\frac{{{r_l}}}{c}},
\end{equation}
\begin{equation}\label{equ:OFDM_signal_3}
	{k_v(m)} = e^{- \frac{j2\pi {\it{m}}T{{v_l}{f_c}}}{c} } .
\end{equation}
Applying DFT for each row of ${{\bf{S}}_g}$, the velocity of target ${v_l}$ can be deduced as follows
\begin{equation}\label{equ:Doppler}
	\begin{aligned}
	&	I^l_{s,n} = {\mathsf {DFT}} ({{\bf{S}}_g}) \\
	&		v_l \in \left [\frac{ {{I^l_{s,n}}} \cdot c}{{T {\color{black} M_I} f_c}},\frac{ \left({{I^l_{s,n}} + 1}\right)  \cdot c}{{T {\color{black} M_I} f_c}} \right )
	\end{aligned},
\end{equation}
where ${I^l_{s,n}}$ is the index of the peak of the DFT of the ${\it{n}}$-th row of ${{\bf{S}}_g}$, {\color{black} $M_I$ is the number of DFT points}.
Applying inverse DFT (IDFT) for each column of ${{\bf{S}}_g}$, the range of target ${r_l}$ can be deduced as follows
\begin{equation}\label{equ:distance}
	\begin{aligned}
		&	I^l_{s,m} = {\mathsf {IDFT}} \left({{\bf{S}}_g}\right) \\
		&	r_l \in \left [\frac{{{I^l_{s,m}} \cdot c}}{{{{\color{black} N_I}  \Delta f}}},\frac{{ \left({I^l_{s,m}} + 1\right) \cdot c}}{{{{\color{black} N_I}  \Delta f}}} \right )
	\end{aligned},
\end{equation}
where ${I^l_{s,m}}$ is the index of the peak of the IDFT of the ${{m}}$-th column of ${{\bf{S}}_g}$, {\color{black} $N_I$ is the number of IDFT points}.
Details of this method is described in \cite{[OFDM]}.

\subsection{${\mathcal F}_3$: ISAC Channel Reconstruction Algorithm}\label{sec:ISAC-NET-1-3}

Since the range and velocity of the $l$-th target $r_l$ and $v_l$ can be obtained by passive sensing, we can reconstruct the ISAC channel ${\mathsf {CHA}} (\cdot)$ based on \eqref{equ:ISAC_new_4_1}, \eqref{equ:ISAC_new_4_2}, \eqref{equ:ISAC_new_4_3}, \eqref{equ:ISAC_new_4_4}, \eqref{equ:ISAC_NET_3-1-1} and \eqref{equ:ISAC_NET_3-1-2}.
Details of the ISAC channel reconstruction algorithm is presented in $\mathrm{Algorithm\ \ref{alg:ISAC_C}}$.
\begin{algorithm} 
	\caption{ISAC channel reconstruction algorithm}  
	\label{alg:ISAC_C}  
	\begin{algorithmic} 
		\STATE \quad 1). Calculate the AoD and AoA of targets by AoD and AoA estimation algorithm \cite{[AOA_9]}.
		\STATE \quad 2). Calculate the delay ${\boldsymbol \tau}$ and doppler frequency shift ${\bf f}_d$: 
		\begin{equation}\label{equ:ISAC_NET_3-1}
			\begin{aligned}
				{\boldsymbol \tau} &= \frac{\bf r}{c}, \\
				{\bf f}_d &= \frac{\bf v}{\lambda} = \frac{ f_c}{c} {\bf v}.
			\end{aligned}
		\end{equation} \\
	    \STATE \quad 3). Reconstruct ISAC channel:
	    \begin{equation}\label{equ:ISAC_NET_3-2}
	     \hat {\bf H}^l = {\mathsf {CHA}} \left( \left[{\bf r}, {\bf v}\right]\right).
	     \end{equation}
	    \STATE \quad \quad 3a). Reconstruct ISAC channel phase information matrix $\hat {\bf H}^l_{\varphi}$ by $\boldsymbol \tau$ and ${\bf f}_d$ based on \eqref{equ:ISAC_new_4_3} and \eqref{equ:ISAC_new_4_4}.   
	    \STATE \quad \quad 3b). Reconstruct ISAC channel amplitude information matrix $\hat {\bf H}^l_{a}$ based on \eqref{equ:ISAC_new_4_2} and \eqref{equ:ISAC_C_2}. 
		\ENSURE Reconstructed ISAC channel matrix $\hat {\bf H}^l$ based on \eqref{equ:ISAC_new_4_1}.
	\end{algorithmic}  
\end{algorithm}

\section{ISAC-NET}\label{sec:ISAC-NET-2}
\begin{figure*}[ht]
	\centering
	\subfigure[\scriptsize{Conventional architecture}.]{\includegraphics[width=0.9\textwidth]{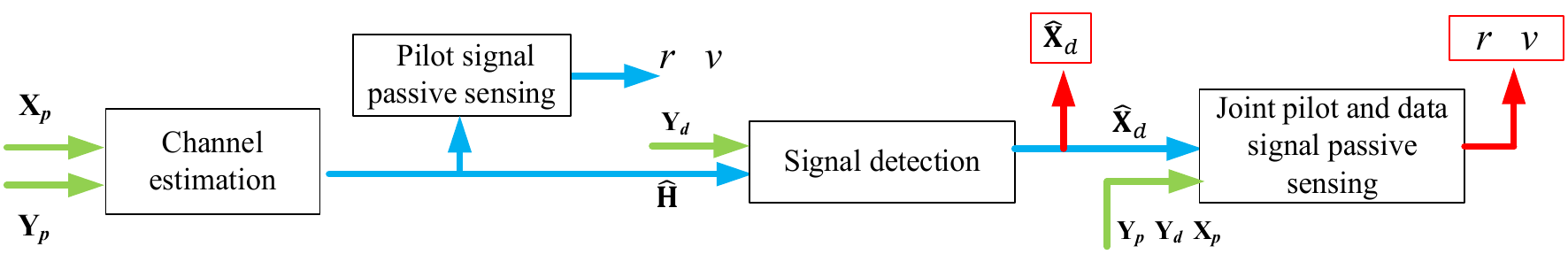}%
		\label{fig:ISAC_NET_0}}
	\hfil
	\subfigure[\scriptsize{ISAC-NET architecture}.]{\includegraphics[width=0.9\textwidth]{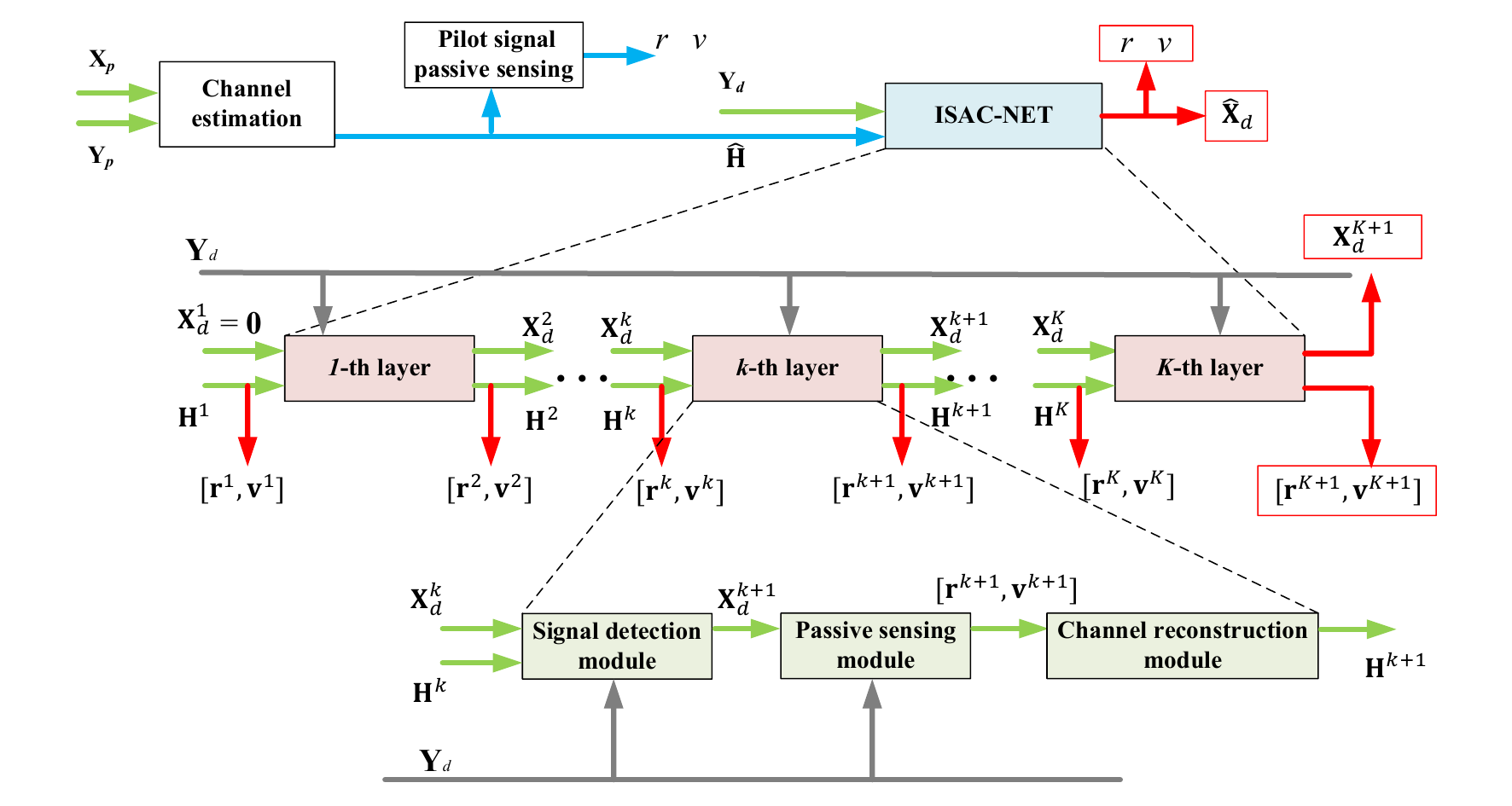}%
		\label{fig:ISAC_NET_1}}
	\hfil
	\caption{Conventional architecture ${\mathcal VS}$ ISAC-NET architecture.}
	\label{fig:ISAC_NET}
\end{figure*}

In the conventional ISAC system, the communication signal detection and passive sensing are carried out in serial. First, the pilot signal is adopted to realize the preliminary passive sensing and channel estimation, and then the signal detection is carried out on the data signal. Then further passive sensing is realized according to the detected data signal. 
This scheme leads to the interference between the error of signal detection and that of passive sensing, which reduces the communication and sensing performance of the system. 

In this subsection, we propose a model-driven ISAC-NET. As $\mathrm{Fig.\ \ref{fig:ISAC_NET}}$ 
\footnote{$\bf H$ represents the channel matrix set ${\bf H}^l$ of $L_p$ paths; $\hat {\bf H}$ represents the estimated channel matrix; $\hat {\bf X}_d$ represents the estimated transmitted signal matrix;} 
shows, there are two parts in the ISAC-NET architecture:
\begin{itemize}
	\color{black}
	\item Pilot signal processing: In this part, since both the transmitted pilot signal and the received pilot signal are known, the traditional channel estimation method can be used to obtain the pilot channel information, $\bf \hat H$. Based on the pilot channel information, the traditional passive sensing algorithm, 2D-DFT, can be used to complete the passive sensing and obtain the rough ranging and velocity measurement results, $\left[\bf r, \bf v\right]$.
	\item ISAC-NET: In this part, we adopt the ISAC signal processing method based on model-driven DL. Compared with the traditional receiver architecture where the signal detector and passive sensing are designed separately, the ISAC-NET considers the characteristics of passive sensing errors in addition to the channel statistics when carrying out signal detection, and uses the detected data signals for passive sensing.
\end{itemize}
In comparison to the data-driven DL-based joint channel estimation and signal detection architecture proposed in \cite{[Main_16],[COM_4]}, which uses a large number of data to train the black-box-based network, we construct the network architecture by employing model-driven DL.
Different with the model-driven DL-based joint channel estimation and signal detection architecture proposed in \cite{[Main]}, which only optimizes the performance of communication signal detection without taking into account the performance of passive sensing, the ISAC-NET can optimize both communication and sensing performance to achieve an optimal trade off between the two functions.
The structure of the ISAC-NET is illustrated in $\mathrm{Fig.\ \ref{fig:ISAC_NET_1}}$, which consists of $K$ cascade layers and each has the same architecture but different trainable parameters.
Each layer of ISAC-NET contains three main modules, the signal detection module, the passive sensing module and the channel reconstruction module, which will be introduced in the following subsection.

\begin{breakablealgorithm} 
	\caption{ISAC signal processing algorithm}  
	\label{alg:CI_ISAC}  
	\begin{algorithmic} 
		\REQUIRE 
		Received data signal ${\bf Y}_d$, 
		estimated channel matrix ${{{\bf H}^l}^{k}}$
		and
		estimated data  signal $ {\bf X}_d^{k}$.
		\STATE \quad 1). Initialize: {\color{black} ${\bf Z}_{d}^1 = {\bf 0}$, ${\bf X}_{d}^1 = {\bf 0}$}.
		\STATE \quad 2). Update ${\bf X}_{d}^{k+1}$: 
		\begin{equation}\label{equ:ISAC_NET_layer_1}
			{\bf X}_{d}^{k+1} = \eta_1^k {\bf X}_{d}^{k} + \sum_{l=0}^{L_p -1} \left({\hat {{{\bf H}^l_a}^k}}\right)^T {\bf Z}_{d}^{k} \left({\hat {{ {\bf H}^l_{\varphi}}^k}}\right)^T.
		\end{equation} \\
		\STATE \quad 3). Update ${\bf Z}_{d}^{k+1}$: 
		\begin{equation}\label{equ:ISAC_NET_layer_2}
			{\bf Z}_{d}^{k+1} =  \eta_2^k {\bf Z}_{d}^{k} + {\bf Y}_d - \sum_{l=0}^{L_p -1} \hat {{\bf H}^l_a}^k {\bf X}_{d}^{k} \hat {{\bf H}^l_{\varphi}}^k.
		\end{equation} \\
		\WHILE{TRUE}
		\STATE {\color{black} $\left[{\bf r}^{k+1}, {\bf v}^{k+1} \right]  = \left[ \left[\quad\right],\left[\quad\right]\right]$}.
		\STATE ${\bf H}^l_{new}$ = ${{\bf H}^l}^k = \left \{ {{\bf H}^l}_a^k , {{\bf H}^l}_{\varphi}^k \right \}$.
		\STATE 4). Passive sensing by 2D-DFT: 
		\begin{equation}\label{equ:ISAC_NET_layer_3}
			\left[M_v, i_v \right] = {\mathsf {DFT}} \left({\bf X}_{d}^{k+1}, {\bf Y}_{d} \right),
		\end{equation} 
		\begin{equation}\label{equ:ISAC_NET_layer_3}
			\left[M_r, i_r \right] = {\mathsf {IDFT}} \left({\bf X}_{d}^{k+1}, {\bf Y}_{d} \right),
		\end{equation} 
		\STATE where $M_v$ and $M_r$ are the maximum DFT and IDFT outputs, {\color{black} $i_v$ and $i_r$ are the index of the maximum DFT and IDFT outputs}.
		\IF {$M_v \ge \eta_3^k$ or $M_r \ge \eta_3^k$}
		\STATE Obtain $v_{tmp}$ and $r_{tmp}$ based on \eqref{equ:Doppler} and \eqref{equ:distance}.
		\STATE ${\bf r}^{k+1} = \left[{\bf r}^{k+1}, r_{tmp} \right]$.
		\STATE ${\bf v}^{k+1} = \left[{\bf v}^{k+1}, v_{tmp} \right]$.
		\STATE Reconstruct ISAC channel: 
		\begin{equation}\label{equ:ISAC_NET_layer_4}
			{\bf H}^l_{tmp} = {\mathsf {CHA}} \left(r_{tmp}, v_{tmp} \right),
		\end{equation} 
		\begin{equation}\label{equ:ISAC_NET_layer_4}
			{\bf H}^l_{new} = {{\bf H}^l}^k - {\bf H}^l_{tmp}.
		\end{equation} \\
		\ELSE
		\STATE break.
		\ENDIF
		\ENDWHILE
		\STATE 5). Update $ \left[{\bf r}^{k+1}, {\bf v}^{k+1} \right]$: 
		\begin{equation} \left[{\bf r}^{k+1}, {\bf v}^{k+1} \right] = \eta_4^k \left[{\bf r}^{k}, {\bf v}^{k} \right] + \left[{\bf r}^{k+1}, {\bf v}^{k+1} \right]
		\end{equation} \\
		\STATE 6). Reconstruct ISAC channel: 
		\begin{equation}\label{equ:ISAC_NET_layer_5}
			{{\bf H}^l}^{k+1} = {\mathsf {CHA}} \left([{\bf r}^{k+1}, {\bf v}^{k+1} ] \right).
		\end{equation} \\
		\STATE 7). Update ${{\bf H}^l}^{k+1}$: 
		\begin{equation}\label{equ:ISAC_NET_layer_6}
			{{\bf H}^l}^{k+1} = \eta_5^k {{\bf H}^l}^{k} + {{\bf H}^l}^{k+1}.
		\end{equation} \\
		\ENSURE Recovered data signal ${\bf X}_{d}^{K+1}$, estimated range and velocity of targets $\left[{\bf r}^{K+1}, {\bf v}^{K+1} \right]$.
	\end{algorithmic}  
\end{breakablealgorithm}

\subsection{Each Layer of ISAC-NET}\label{sec:ISAC-NET-2-1}

\begin{figure*}[ht]
	\includegraphics[scale=0.5]{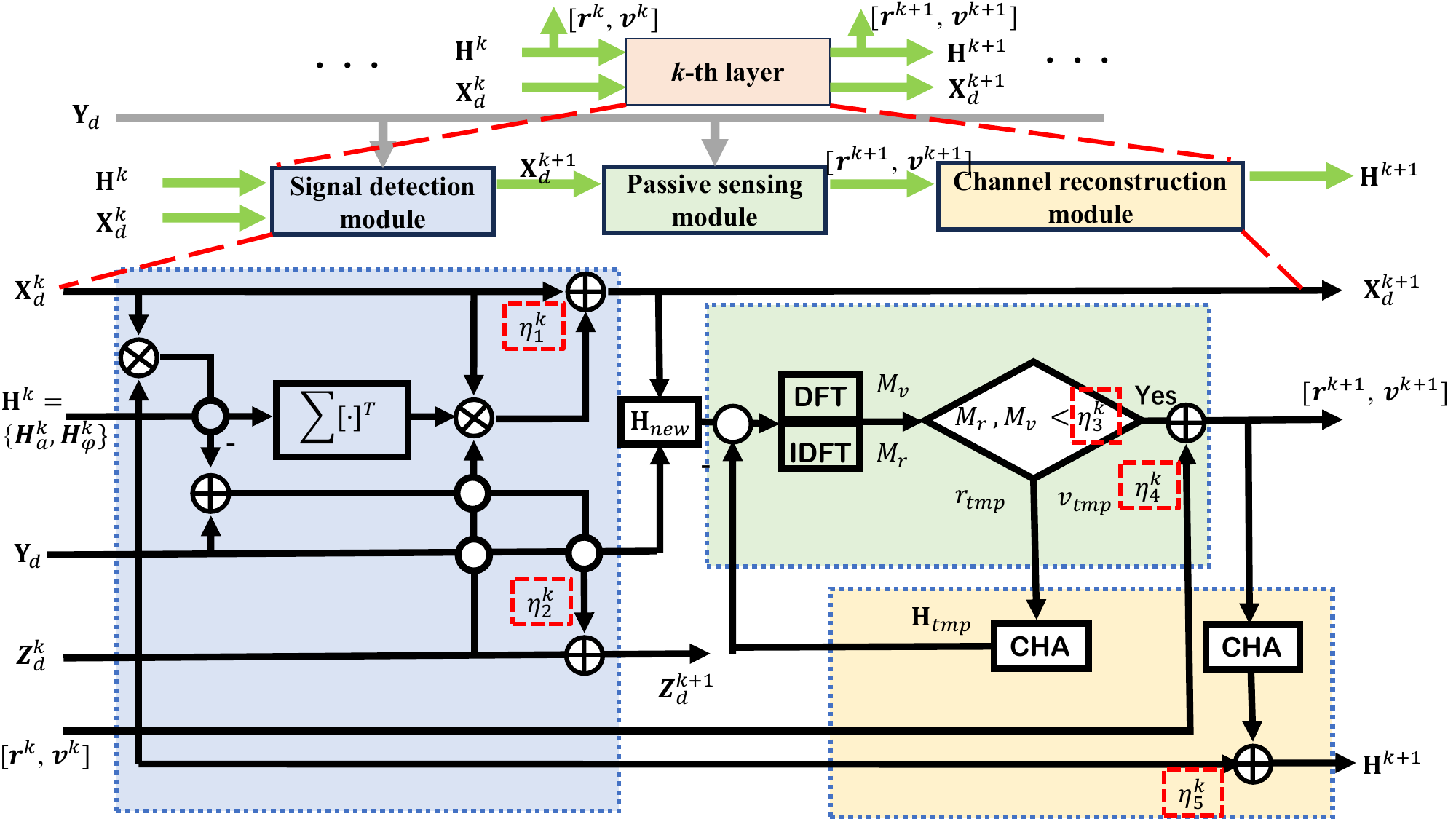}
	\centering
	\caption{Each layer of ISAC-NET.}
	\label{fig:ISAC_NET_layer}
\end{figure*}

As $\mathrm{Fig.\ \ref{fig:ISAC_NET_layer}}$ 
shows, for the $k$-th layer of the ISAC-NET, 
{\color{black}
\begin{itemize}
	\item the input is the estimated data signal $ {\bf X}_d^{k}$, the received data signal ${\bf Y}_d$, the estimated channel matrix ${\bf H}^k$ and the noise matrix ${\bf Z}^{k}$ from the $(k-1)$-th layer. 
	\item The ISAC signal processing module of each layer can be performed as in $\mathrm{Algorithm\ \ref{alg:CI_ISAC}}$, which is developed by unfolding the improved D-AMP algorithm, 2D-DFT algorithm and ISAC channel reconstruction algorithm.  
	\item The blue part of $\mathrm{Fig.\ \ref{fig:ISAC_NET_layer}}$ denotes the improved D-AMP algorithm, the green part denotes the 2D-DFT passive sensing algorithm, the yellow part denotes the ISAC channel reconstruction algorithm.
\end{itemize}
}
Compared with the conventional 2D-DFT passive sensing algorithm, the ISAC signal processing algorithm realizes the multi-target sensing by alternating between the estimation of target parameters and channel reconstruction.
This method has certain improvement in the performance of sensing to multiple targets, which is verified in the subsequent simulation analysis.
Moreover, there are five main learnable variables $\Omega^k = \left \{ \eta_1^k , \eta_2^k, \eta_3^k, \eta_4^k, \eta_5^k   \right \}$ in each layer, which is much less than the data-driven DL, greatly reducing the complexity of network training.

\subsection{Practical Implementation}\label{sec:ISAC-NET-2-2}

The developed ISAC-NET can be divided into the following stages. 
\begin{itemize}
	{\color{black}
	\item \textbf{Data generation:} 
	Inputs of the ISAC-NET is the received signal, Outputs of the ISAC-NET is the demodulated transmitted signal and estimated target velocity and target distance.
	We obtained training and testing data sets through MATLAB simulation, in which $95 \%$ data is used for training and $5 \%$ data is used for testing. The data set includes the transmitted signal set and the received signal set under different signal-to-noise ratio (SNR) conditions, from $-20 \sim 60$ dB. 
	The transmitted signal is obtained by modulating randomly generated bit data.
	The transmitted signal generates the received signal through the ISAC channel, which includes target parameters such as target velocity and target distance.
	\item \textbf{Training settings:} 
	Each batch of training data is set up as an OFDM signal matrix containing 1024 subcarriers and 256 OFDM symbols. The batch size of each training epoch is 128, and the number of training epochs is 500. The learning rate is set to 0.001.
	}
	\item \textbf{Offline training stage:} 
	In the offline training stage, we obtain the optimized parameters, ${\bf \Omega} = \left \{\Omega^k \right \}_{k=1}^K$ for different SNRs based on the tensorflow platform. 
	\item \textbf{Deployment stage:} 
	The optimized parameters are stored to detect the
	modulated symbols in the deployment stage. 
	The ISAC-NET can be interpreted as a new iterative detector after training. 
	The incorporated learnable parameters can adapt to practical channels, compensate for signal detection errors and passive sensing errors, and improve communication and sensing performance.
\end{itemize}
Although aforementioned implementation process is in an offline manner, the proposed ISAC-NET can also be implemented by online training to adapt to the fluctuations in the channel conditions owing to the superiority of its low demand for training data and computational resources.

\section{Simulation Results}\label{sec:Simulation}
{\color{black}
To demonstrate the relationship between the communication demodulation error and sensing accuracy in the ISAC system,
we first analyze the communication performance with sensing errors and the sensing performance in the condition of communication demodulation errors. 
To verify the improvement of ISAC-NET in the communication and sensing performance, the communication and sensing performance of the proposed ISAC-NET is simulated and compared with the typical communication and sensing signal processing algorithms.
It should be noted that, the SNR adopted in the simulation is denoted the SNR of the received signal, and each simulation in this paper is calculated over 500 Monte Carlo trials.
Simulation parameters used in this section are shown in $\mathrm{TABLE\ \ref{Parameter:simulation}}$ \cite{[5G_signal],[Main]}.
}

\begin{table}[h]
	\caption{Simulation parameters adopted in this paper.}
	\centering
	\label{Parameter:simulation}
	\begin{tabular}{c|c|l|c|c|l}
		\hline
		\hline
		Items & Value & Meaning of the parameter & Items & Value & Meaning of the parameter \\ \hline
		$f_c$ & 4 GHz & Carrier frequency & $\Delta f$ & $120$ kHz & Carrier frequency \\ \hline
		 $M$ & 256 & Number of OFDM symbols & $N$ & 1024 & Number of subcarriers \\ \hline
		\color{black} $M_I$ & \color{black} 2560 & \color{black} Number of DFT points & \color{black} $N$ & \color{black} 10240 &\color{black}  Number of IDFT points \\ \hline
		$T_p$ & 8.3 $\mu$s & OFDM symbol period & $T_c$ & 2.08 $\mu$s & CP period \\ \hline
		$T$ & 10.38 $\mu$s & The whole OFDM period & $B$ & 123 MHz & Frequency bandwidth \\ \hline
		$N_t$ & 8  & Number of transmitting antenna array & $N_r$ & 8  & Number of receiving antenna array \\ \hline
		$L$ & 1 $\sim$ 3  & Number of targets & $\sigma$ & 1 m$^2$ & RCS of targets \\ \hline
		16 QAM &   & Modulation type & SNR & $-20 \sim 60$ dB  & SNR of received signals \\ \hline
		$w_p$ & 1/14  & Pilot signal length ratio & &   &  \\ \hline
	\end{tabular}
\end{table}

\subsection{Communication and Sensing Performance Under Error Conditions}\label{sec:Simulation-1}

\begin{figure}[ht]
	\centering
	\subfigure[\scriptsize{NMSE of range}.]{\includegraphics[width=0.48\textwidth]{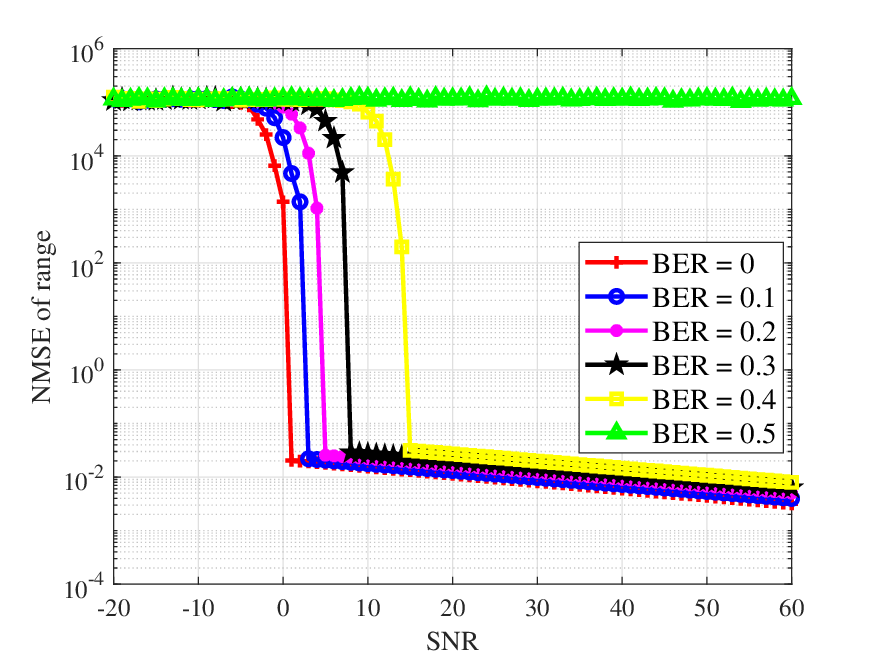}%
		\label{fig:RMSE_R_BER}}
	\hfil
	\subfigure[\scriptsize{NMSE of velocity}.]{\includegraphics[width=0.48\textwidth]{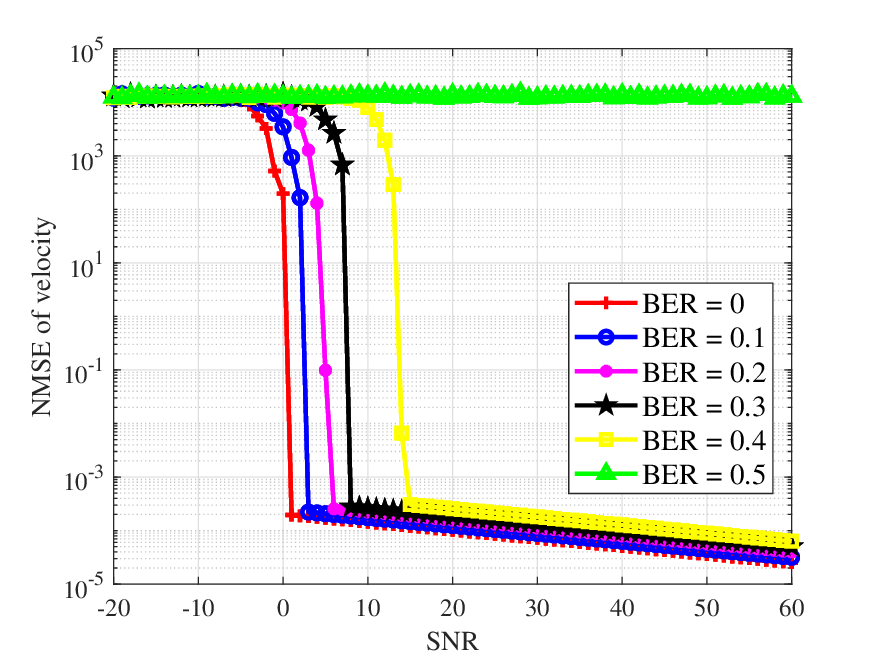}%
		\label{fig:RMSE_V_BER}}
	\hfil
	\caption{RMSE of range and velocity with different BERs.}
	\label{fig:RMSE_BER}
\end{figure}

\begin{figure}[ht]
	\centering
	\subfigure[\scriptsize{BER}.]{\includegraphics[width=0.48\textwidth]{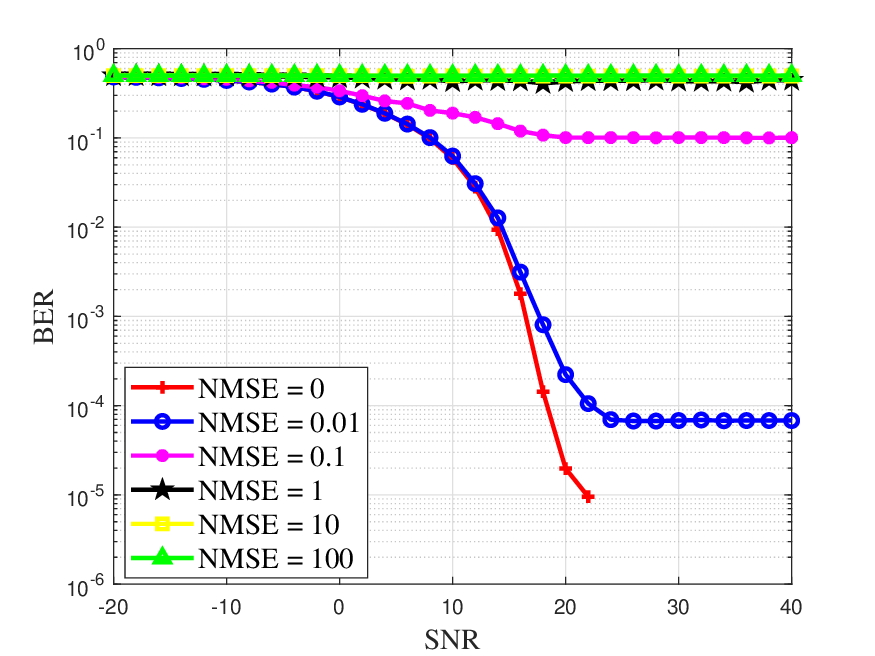}%
		\label{fig:BER_RMSE}}
	\hfil
	\subfigure[\scriptsize{SER}.]{\includegraphics[width=0.48\textwidth]{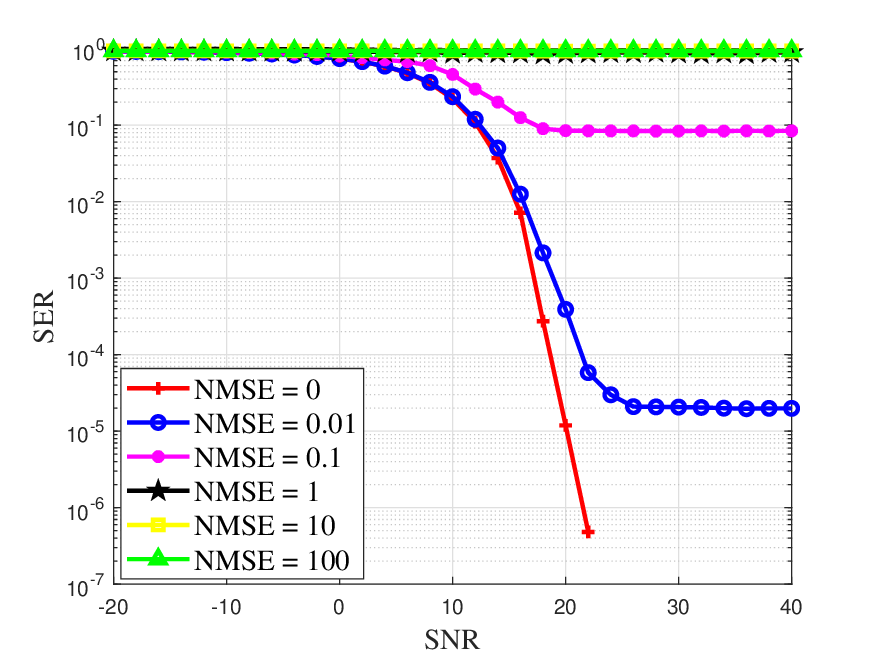}%
		\label{fig:SER_RMSE}}
	\hfil
	\caption{BER and SER with different RMSEs.}
	\label{fig:SER_BER_RMSE}
\end{figure}

In this section, we analyze and simulate the communication and sensing performance in the condition of communication demodulation errors. 
{\color{black}
$\mathrm{Fig.\ \ref{fig:RMSE_BER}}$ presents the normalized mean squared errors (NMSEs) of range and velocity with different SNRs, which can be expressed as 
\begin{equation} \label{equ:Simulation_1_1}
	\begin{aligned}
		\Gamma_r &= \frac{1}{L} \sum_{l=0}^{L-1} \frac{\left |\hat r_{l} - r_l \right|^2}{r_l^2}, \\
		\Gamma_v &= \frac{1}{L} \sum_{l=0}^{L-1} \frac{\left|\hat v_{l} - v_l\right|^2}{v_l^2}
	\end{aligned},
\end{equation}
where $L$ is the number of targets, $r_l$ and $v_l$ are the actual range and velocity of the $l$-th target, $\hat r_{l}$ and $\hat v_{l}$ are the estimated range and velocity of the $l$-th target.
}
$\mathrm{Fig.\ \ref{fig:SER_BER_RMSE}}$ presents the bit error rate (BER) and symbol error rate (SER).

{\color{black}
As $\mathrm{Fig.\ \ref{fig:RMSE_R_BER}}$ and $\mathrm{Fig.\ \ref{fig:RMSE_V_BER}}$ show that, under the same BER condition, the NMSEs of range and velocity decrease with the increase of SNR, which means that the passive sensing accuracy is improved.
In addition, under the same SNR condition, the NMSEs of range and velocity increase with the increase of BER, which means that the passive sensing accuracy is decreased.
When the BER reaches 0.5, it indicates that the recovered communication signal is completely wrong. At this time, the NMSEs of range and velocity do not decrease with the increase of SNR, and the sensing accuracy can no longer decrease.
Therefore, we conclude that the accurate demodulation of the communication symbols is one of the preconditions for obtaining high-accurate passive sensing performance.
}

As $\mathrm{Fig.\ \ref{fig:BER_RMSE}}$ and $\mathrm{Fig.\ \ref{fig:SER_RMSE}}$ show that, under the same NMSE condition, with the increase of SNR, the BER and SER decrease, which means that the accuracy of communication demodulation is improved.
In addition, under the same SNR condition, the BER and SER increase with the increase of NMSE, which means that the accuracy of communication demodulation is decreased.
{\color{black}
When the NMSE reaches 100, the BER is close to 0.5, which indicates that the recovered communication signal is completely wrong. 
Therefore, we conclude that accurate passive sensing is one of the preconditions for obtaining high-accurate communication performance.

The simulation results provide a theoretical basis for the proposed ISAC-NET based on communication and sensing iterative optimization.
}

\begin{figure}[h]
	\centering
	\subfigure[\scriptsize{BER}.]{\includegraphics[width=0.48\textwidth]{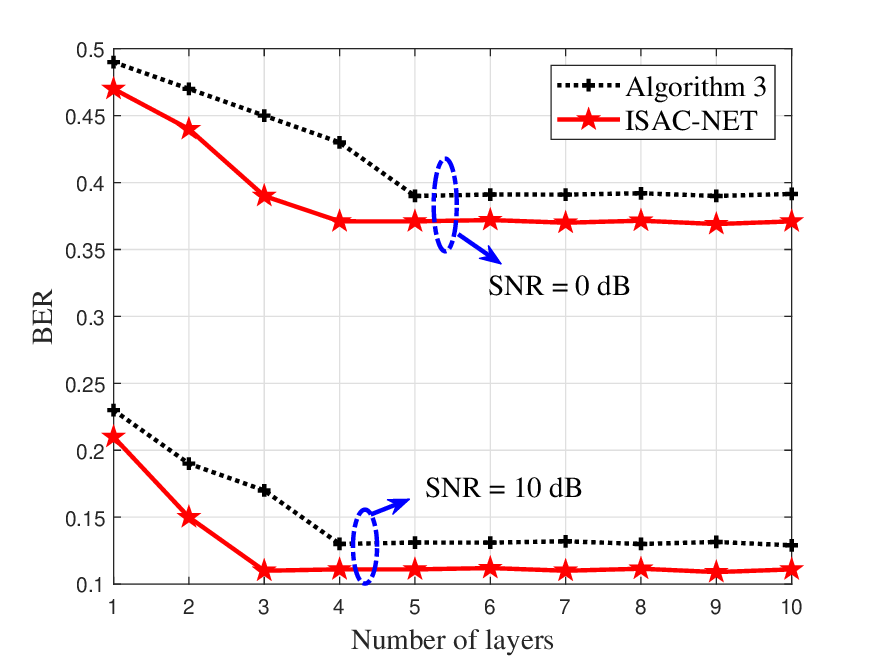}%
		\label{fig:Conver_BER}}
	\hfil
	\subfigure[\scriptsize{NMSE of range}.]{\includegraphics[width=0.48\textwidth]{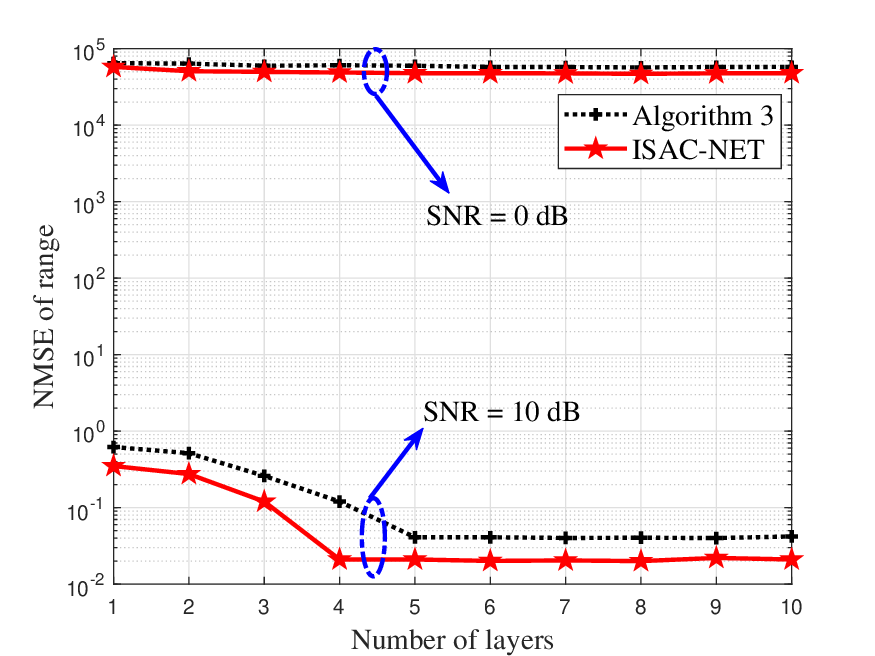}%
		\label{fig:Conver_NMSE_R}}
	\hfil
	\caption{BER and NMSE of the ISAC-NET and $\mathrm{Algorithm\ \ref{alg:CI_ISAC}}$ versus the
		number of layers with different SNRs}
	\label{fig:Convergence}
\end{figure}

\begin{figure}[h]
	\centering
	\subfigure[\scriptsize{BER}.]{\includegraphics[width=0.48\textwidth]{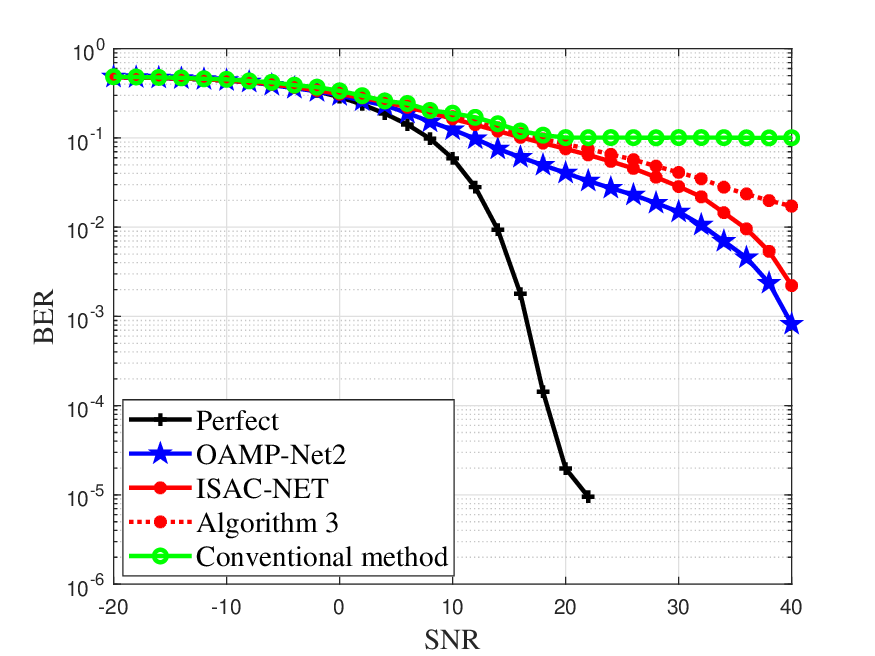}%
		\label{fig:BER_ISAC_NET_1}}
	\hfil
	\subfigure[\scriptsize{SER}.]{\includegraphics[width=0.48\textwidth]{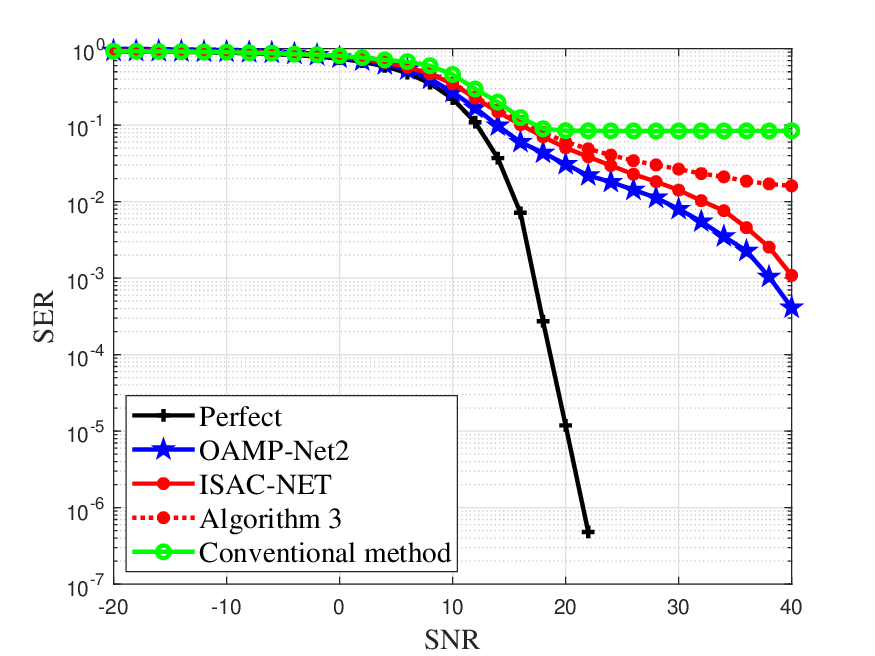}%
		\label{fig:SER_ISAC_NET_1}}
	\hfil
	\caption{BER and SER comparison of the ISAC-NET with other typical algorithms.}
	\label{fig:BER_SER_ISAC_NET_1}
\end{figure}

\begin{figure}[h]
	\centering
	\subfigure[\scriptsize{NMSE of range}.]{\includegraphics[width=0.48\textwidth]{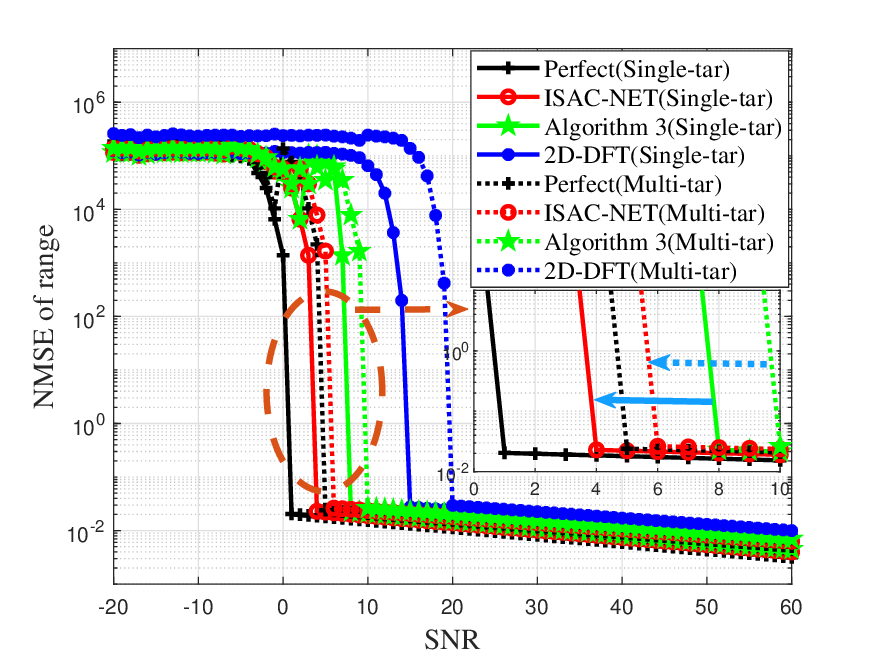}%
		\label{fig:NMSE_R_ISAC_NET_1}}
	\hfil
	\subfigure[\scriptsize{NMSE of velocity}.]{\includegraphics[width=0.48\textwidth]{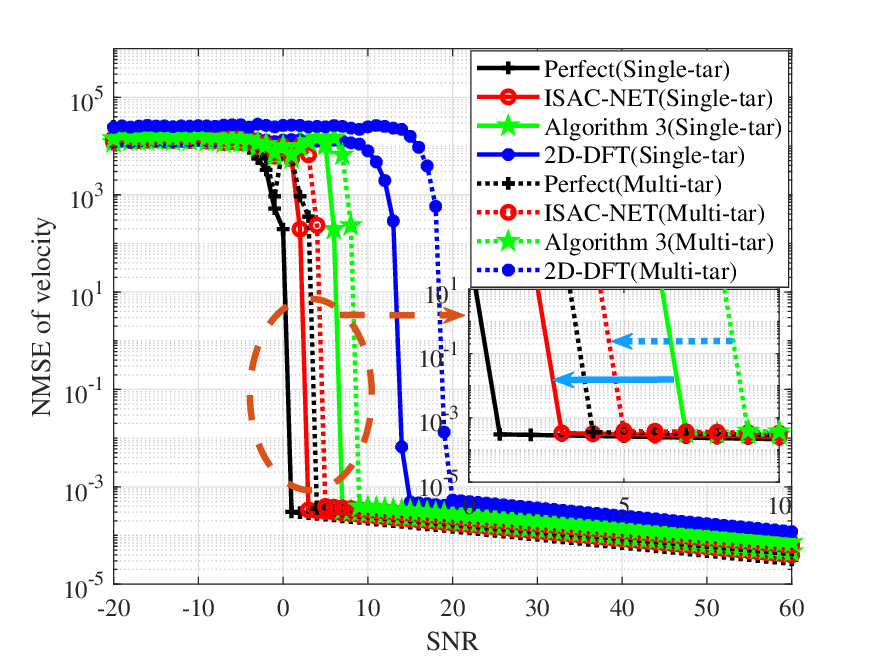}%
		\label{fig:NMSE_V_ISAC_NET_1}}
	\hfil
	\caption{NMSE comparison of the ISAC-NET with other typical algorithms.}
	\label{fig:NMSE_R_V_ISAC_NET_1}
\end{figure}

\subsection{Communication and Sensing Performance of ISAC-NET}\label{sec:Simulation-3}
{\color{black}
In this subsection, we analyze and simulate the communication and passive sensing performance of ISAC-NET. 
Since convergence is a necessary condition of the iterative algorithm, we first simulate and verify the convergence of ISAC-NET in sensing and communication.
After that, the communication and sensing performance of the proposed ISAC-NET is simulated and compared with the typical communication and sensing signal processing algorithms.
}
\subsubsection{Convergence Analysis}\label{sec:Simulation-3-1}
First, we analyze the convergence performance of $\mathrm{Algorithm\ \ref{alg:CI_ISAC}}$ and ISAC-NET. 
$\mathrm{Fig.\ \ref{fig:Convergence}}$ illustrates the BER and NMSE versus the number of layers (iterations) under $\mathrm{SNR\ =\ 0\ dB}$ and $\mathrm{SNR\ =\ 10\ dB}$.
From $\mathrm{Fig.\ \ref{fig:Conver_BER}}$, both the $\mathrm{Algorithm\ \ref{alg:CI_ISAC}}$ and
ISAC-NET converge within five layers (iterations) for all
the cases.
From $\mathrm{Fig.\ \ref{fig:Conver_NMSE_R}}$, both the $\mathrm{Algorithm\ \ref{alg:CI_ISAC}}$ and ISAC-NET converge within five layers (iterations) for all
the cases when $\mathrm{SNR\ =\ 10\ dB}$. 
{\color {black}
Moreover, the estimation error of ISAC-NET is smaller than that of $\mathrm{Algorithm\ \ref{alg:CI_ISAC}}$. 
}
Since $\mathrm{SNR\ =\ 0\ dB}$ does not meet the requirement of passive sensing, it is difficult to achieve high-accurate target detection, both the $\mathrm{Algorithm\ \ref{alg:CI_ISAC}}$ and ISAC-NET cannot converge.

\begin{table*}[ht]
	\centering
	\caption{Communication and sensing algorithms for simulation analysis}
	\label{label:algorithm}
	\begin{tabular}{l|l|l}
		\hline \hline
		Method & Communication & Sensing \\ \hline
		Perfect & \begin{tabular}[c]{@{}l@{}}MAP without CSI error\end{tabular}
		& 2D-DFT \cite{[OFDM]} without communication demodulation error  \\ \hline
		ISAC-NET & ISAC-NET with CSI error & \begin{tabular}[c]{@{}l@{}}ISAC-NET with communication demodulation error\end{tabular}
		\\ \hline
		OAMP-Net2 & OAMP-Net2 proposed in \cite{[Main]} with CSI error &  \\ \hline
		Conventional method & MAP with CSI error &  \\ \hline
		2D-DFT &  & 2D-DFT \cite{[OFDM]} with communication demodulation error \\ \hline
	\end{tabular}
\end{table*}

\subsubsection{Performance Comparison}\label{sec:Simulation-3-2}

{\color {black}
Through simulations, the communication and sensing performance of the proposed ISAC-NET is evaluated and compared to the performance of typical communication and sensing signal processing algorithms, which are listed in $\mathrm{TABLE\ \ref{label:algorithm}}$.
For the multi-targets detection scenario, the number of targets need to detect is set as three. 
The actual range and velocity of targets are $10\ \rm{m}$, $50\ \rm{m}$, $100\ \rm{m}$ and $5\ \rm{m/s}$, $10\ \rm{m/s}$, $15\ \rm{m/s}$ respectively.
}

$\mathrm{Fig.\ \ref{fig:BER_SER_ISAC_NET_1}}$ compares the BER and SER of MAP \cite{[Main]}, OAMP-Net2 \cite{[Main]}, Algorithm \ref{alg:CI_ISAC}, and ISAC-NET. 
The perfect line denotes the BER and SER without CSI. 
The conventional method refers to the direct demodulation algorithm of the received signal, the MAP algorithm, with channel estimation error.
{\color{black}
As $\mathrm{Fig.\ \ref{fig:BER_SER_ISAC_NET_1}}$ shows, ISAC-NET and OAMP-Net2 have similar BER and SER, smaller than Algorithm \ref{alg:CI_ISAC}, which can verify the communication performance gain of deep unfolding over conventional algorithm without unfolding. 
}

$\mathrm{Fig.\ \ref{fig:NMSE_R_V_ISAC_NET_1}}$ compares the NMSE of the 2D-DFT algorithm \cite{[OFDM]} and ISAC-NET. 
The perfect line denotes the NMSE without communication demodulation errors. 
As $\mathrm{Fig.\ \ref{fig:NMSE_R_V_ISAC_NET_1}}$ shows, under the same SNR condition, NMSE of range and velocity by ISAC-NET is smaller than 2D-DFT. 
It means that ISAC-NET can obtain high-accurate passive sensing performance.
{\color{black}
Moreover, compared with the ISAC signal processing algorithm without unfolding (Algorithm \ref{alg:CI_ISAC}), ISAC-NET can obtain smaller NMSE of range and velocity, which can verify the sensing performance gain of deep unfolding over conventional algorithm without unfolding. 
}

\section{Conclusion}\label{sec:Conclusion}

In this paper, we proposed an integrated sensing and communication network based on model-driven DL, called ISAC-NET, that combines passive sensing with communication demodulation signal processing using model-driven DL.
{\color {black}
Dissimilar to existing passive sensing algorithm that the communication signal detection and passive sensing are carried out in serial.
The ISAC-NET could obtain passive sensing results and communication demodulated symbols simultaneously, 
which adopted the block-by-block signal processing method. The ISAC-NET was divided into the passive sensing module, signal detection module and channel reconstruction module.
}
Moreover, the communication and sensing performance improvements of ISAC-NET in range and velocity estimation, BER and SER are analyzed and simulated.
The simulation results verified that the ISAC-NET could obtain better communication and passive sensing performance.

\begin{appendices}

\section{Doppler frequency shift for passive sensing} \label{app:A}

\begin{figure}[ht]
	\includegraphics[scale=0.4]{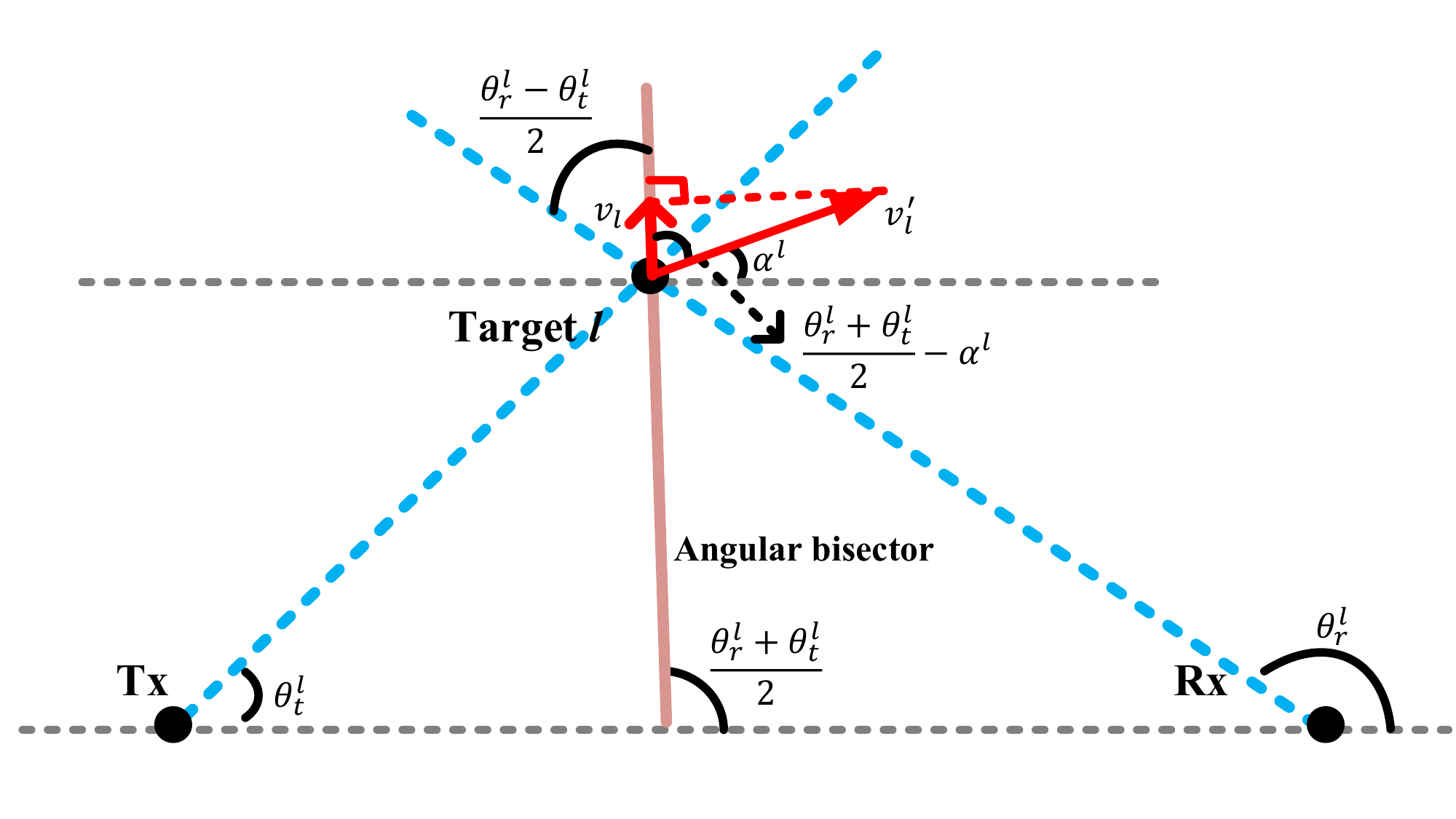}
	\centering
	\caption{Doppler frequency shift for passive sensing.}
	\label{fig:PS_vs_AS_2}
\end{figure}
As $\mathrm{Fig.\ \ref{fig:PS_vs_AS_2}}$, the $l$-th target is moving with velocity ${v_l}'$ in the direction $\alpha^l$. The frequency of the ISAC signal at $\rm Tx$ is $f_c$.
Based on the source reference frame of relativistic Doppler effects \cite{[DP]}, the frequency of the ISAC signal arriving at the $l$-th target can be expressed as 
\begin{equation}\label{equ:PS_vs_AS_2_3}
	\begin{aligned}
		f_1  &= f_c \frac{1-\beta {\rm cos} \left(\alpha^l - \theta_t^l \right)}{ \sqrt {1 - \beta^2}}
	\end{aligned},
\end{equation}
where $\beta = \frac {{v_l}'}{c} \ll 1$ with $c$ being the velocity of light.
Similarly, the frequency of the ISAC signal at $\rm Rx$ can be given by
\begin{equation}\label{equ:PS_vs_AS_2_4}
	\begin{aligned}
		f_2  &= f_1 \frac{ \sqrt {1 - \beta^2}}{1+\beta {\rm cos} \left(\alpha^l - \theta_r^l \right)}
	\end{aligned}.
\end{equation}
The doppler frequency shift of the $l$-th target can be expressed as
\begin{equation}\label{equ:PS_vs_AS_2_5}
	\begin{aligned}
		f_{d,l} &= f_2 - f_c \\
		  &= f_c \left(  { \frac{ \sqrt {1 - \beta^2}}{1+\beta {\rm cos} \left(\alpha^l - \theta_r^l \right )}  \cdot  \frac{1-\beta {\rm cos} \left(\alpha^l - \theta_t^l \right)}{ \sqrt {1 - \beta^2}} }           -1 \right) \\
		  &= f_c \left( \frac{1-\beta {\rm cos} \left(\alpha^l - \theta_t^l \right)}{1+\beta {\rm cos} \left(\alpha^l - \theta_r^l \right)}     -1 \right) \\
		  &= -\beta f_c \frac{ {\rm cos} \left(\alpha^l - \theta_t^l \right) +  {\rm cos} \left(\alpha^l - \theta_r^l \right) }{1+\beta {\rm cos} \left(\alpha^l - \theta_r^l \right)}	 
	\end{aligned}.
\end{equation}
Based on $\beta = \frac {{v_l}'}{c} \ll 1$, the doppler frequency shift of the $l$-th target $f_{d,l}$ can be approximated as
\begin{equation}\label{equ:PS_vs_AS_2_6}
	\begin{aligned}
		f_{d,l} &\approx -\frac {{v_l}'}{c} f_c \cdot {\rm cos} \left(\frac{\theta_r^l - \theta_t^l}{2} \right) \cdot {\rm cos} \left(\alpha^l - \frac{\theta_r^l + \theta_t^l}{2} \right)
	\end{aligned}.
\end{equation}

\end{appendices}

\bibliographystyle{IEEEtran} 
\bibliography{reference}

\ifCLASSOPTIONcaptionsoff
  \newpage
\fi

\end{document}